\newcommand{\beq}{  \begin{eqnarray}}
\newcommand{\eeq}{  \end{eqnarray}}

\documentclass[pra,twocolumn,shopacs,preprintnumbers,asmath,amssymb]{revtex4-1}
\usepackage{subfig}
\usepackage{graphicx}
\usepackage{dcolumn}
\usepackage{bm}

\begin{document}
    	
\title{Geometric Phase Atom Optics and Interferometry}
    
\author{B. Zygelman}
\email{bernard@physics.unlv.edu}
\affiliation{%
Department of Physics and Astronomy, University of Nevada, Las Vegas, Las Vegas NV 89154
}%

\date{\today}
\begin{abstract}
We illustrate how geometric gauge forces and topological phase effects emerge in atomic and molecular systems without employing assumptions that rely on adiabaticty.  We show how geometric magnetism may be harnessed to engineer novel quantum devices including a velocity sieve, a component in mass spectrometers, for neutral atoms. We introduce and outline a possible experimental setup that demonstrates topological interferometry for neutral spin 1/2 systems. For that 2-level system, we study the transition from Abelian to non-Abelian behavior and  explore its relation to the molecular Aharonov-Bohm (MAB) effect. 
\end{abstract}

\maketitle
\section{Introduction}
Berry's phase\cite{ber84}, a realization
of a non-integrable phase factor\cite{Dirac31,wuyang75}, plays an important role in describing adiabatic quantum
evolution in a semi-classical setting. It is also known\cite{mead76,moo86,zyg87a,jackiw88,ber89}  that
effective gauge potentials, that give rise to such phases, emerge 
in fully quantal systems in which true adiabaticity is ill-defined.
In addition to generating phase holonomies\cite{mead76},
they may lead to effective Lorentz-like forces\cite{zyg87a,ber89,ber93} 
acting on the quantum system. 

Today, applications of geometric gauge forces in cold many-body systems\cite{dal11,spiel14}
is an active and topical area of research.
Dressing atoms using lasers\cite{juz06}, researchers\cite{lin09a,lin09b} have been able to engineer Lorentz-like forces, 
an effect sometimes called geometric\cite{ber93,zyg12}, synthetic\cite{spiel14} or artificial magnetism\cite{dal11},
in ensembles of cold atoms. It is hoped that the latter may allow realization of novel quantum Hall physics in a quantum degenerate gas.

Another possible application of geometric magnetism is in  the manipulation of individual neutral atoms or neutrons\cite{zyg12,zyg12b,kiffner2}.
In Refs. \cite{zyg12,zyg12b}  we proposed how  gauge forces could be exploited to construct 
``magnetic lenses"  for neutral  matter. In this paper we elaborate on those observations by introducing two novel and striking
illustrations of the latter. In the first example we illustrate 
 how this phenomenon allows realization of a velocity sieve, a component in mass spectrometers, for beams of neutral particles.  In another example we consider a quantum mechanical analog\cite{march92,shen96,zyg12} of a field theoretical model\cite{march92} to demonstrate topological quantum interferometry for
neutral spin-1/2 systems. It's laboratory realization could have applications in topological quantum computing protocols.

With some exceptions e.g. \cite{zyg12,kiffner2,gross15}, most theoretical studies, by necessity, 
have relied  on some form of the adiabatic assumption, i.e. the Born-Oppenheimer (BO)
approximation. Gauge potentials are explicit in the adiabatic picture but it also known\cite{zyg90,mead92} that those  effective (non-Abelian) gauge 
potentials describe a pure gauge\cite{wuyang75}.  

Consider the following Schroedinger equations 
\beq
-\, \frac{\hbar^{2}}{2m} {\bm \nabla_{\bm R} }^{2} \psi + V({\bm R}) \psi = i \hbar \frac{\partial \psi}{\partial t} 
\label{0.1}
\eeq
and 
\beq
-\, \frac{\hbar^{2}}{2m} {({\bm \nabla}_{\bm R}  - i {\bm A}     ) }^{2} \psi' + V({\bm R}) \psi' = i \hbar \frac{\partial \psi'}{\partial t} 
\label{0.2}
\eeq
for amplitudes $\psi$, $\psi'$ respectively. 
We take them to be $n$ dimensional column vectors and $V({\bm R})$ is a local $n \times n$ matrix potential. The gauge potential
\beq
{\bm A } \equiv  i \, U^{\dag}({\bm R}) \, {\bm \nabla}_{\bm R} \, U({\bm R})
\label{0.2a}
\eeq
where $U({\bm R})$ is a differentiable, single-valued, unitary $n \times n$ matrix. According to definition (\ref{0.2a})  ${\bm A}$ is a
pure gauge \cite{wuyang75,zyg90}. 
We pose the question; do the Schroedinger equations (\ref{0.1})  and (\ref{0.2}) describe the same physics?  If $n=1$ the
answer is in the affirmative since then $ U({\bm R}) = \exp(-i \Omega )$ and
\beq  {\bm A} \equiv  i\,  \exp(i\Omega) \,  {\bm \nabla}_{\bm R} \, \exp(-i\Omega) = {\bm \nabla}_{\bm R} \Omega
\label{0.2b}
\eeq
where $ \Omega$ is a single-valued function of $ {\bm R}$. 
 Therefore, $\psi'=\exp(i \Omega) \, \psi$ and the physical content of Eqs (\ref{0.1}) and (\ref{0.2}) are identical\cite{gott}, 
 as the eigenvalues of physical operators, which transform in a covarient manner, are invarient under this gauge transformation. 
 The same is true for the non-Abelian case where $ n > 1$, provided that $ [U,V]=0 $ since then $ \psi'= U({\bm R}) \, \psi$.  
So for these cases we find that the minimal coupling of a pure gauge potential is fully equivalent
to a description where $ {\bm A} =0 $.
However this is not the case  if  $ [{\bm A},V] \neq 0 $, as the amplitudes $\psi$, and $\psi'$ 
are no longer related by the (single-valued) gauge transformation $U$. 
The conclusion follows from the fact that $[U({\bm R}),V] \neq 0$ 
 if the above inequality holds (see also Appendix C). The implications of this observation will become apparent in the discussion below.

Are non-trivial  gauge forces
an artifact of the adiabatic approximation\cite{gross15} ? Are singularities in the adiabatic Hamiltonian solely responsible for the emergence of topological
phases? Several interpretations\cite{aharonov90,aharonov92,gold05} for the origins of geometric gauge forces have been
advanced. However, compelling examples that offer fully quantum solutions to systems in which such forces arise in the adiabatic limit
have largely been unavailable and, therefore, predictions are limited 
by the validity of assumptions based on adiabaticity. 

In order to address questions and deficiencies in theoretical approaches that assume adiabaticity, 
we \cite{zyg12,zyg12b} introduced a wave packet propagation scheme that does not require the assumptions and
restrictions imposed by it. The resulting time dependent solutions for the systems are exact, 
within the bounds imposed by numerical error, and they 
are not compromised by issues relating to the robustness of
the adiabatic approximation. This allows us to make definitive 
verdicts on the fidelity of predictions informed by the latter, 
and the gauge theory interpretation that follows from it.

In the discussion below we first present the general theoretical framework for the specific cases considered. We 
introduce a 2-level system whose adiabatic Hamiltonian is defined in such a way that there is
a cancellation of the effective Lorentz force with the gradient force in low energy solutions to
the coupled Schroedinger equation. This cancellation is similar to that which occurs when appropriate
external electric and magnetic fields ``select'' the velocity of a charged particle beam. This theoretical
model illustrates how geometric magnetism could be harnessed to entertain similar capabilities for neutral  systems.
We then focus our discussion on neutral spin-1/2 systems subjected to an external magnetic field. We show, with the appropriate
external field configuration, how such a system can exhibit topological Aharonov-Bohm\cite{ab59} (AB) behavior. 
Though it has been discussed previously\cite{march92,shen96,zyg12} within the framework of the BO approximation,  here
we offer fully time dependent, coupled channel, wave packet solutions without employing assumptions based on adiabaticity. 
This capability allows us extend the analysis to cases in which the BO approximation is no longer valid. We investigate
how this systems transitions at low energies, and one that allows an Abelian description in which the AB gauge vector potential emerges, 
to that in which non-Abelian features arise. Finally, we explore the relationship of this system to a phenomenon that occurs
in molecular systems that possess conical intersections\cite{teller,Herzberg,mead76,gross15}. Atomic units will be used throughout, unless otherwise indicated.   
\section{Theory}
\subsection{Velocity sieve}
Consider the model Hamiltonian for an atom, or n-state, system
\beq
&& H= -\, \frac{\hbar^{2}}{2 m} {\bm \nabla_{R}}^{2} + H_{ad}({\bm R}). 
\label{z0.1a}
\eeq
$ {\bm R}$ is the position operator for the atom and $H_{ad}({\bm R})$ is the adiabatic Hamiltonian that
generates its internal dynamics. Suppose it can be decomposed into
the form $U({\bm R}) H_{BO} U^{\dag} ({\bm R}) $, where $ H_{BO}({\bm R})$ is an n-dimensional  diagonal matrix with eigenvalues 
$ \epsilon_{i}({\bm R})$, and
$ U({\bm R})$ is a unitary matrix. In this discussion we consider the case where $n=2$, and $ {\bm R} =(x,y)$ and limit $H_{ad}({\bm R})$
to be time independent.
We require $U({\bm R})$ to be single-valued\cite{comment} and  express it
 \beq
 && U = \exp(-i \sigma_{3} \phi/2) \exp(-i \sigma_{2} \Omega/2 ) \exp(i \sigma_{3} \phi/2) 
 \label{S1.1}
\eeq
 where $\sigma_{i}$ are the Pauli matrices, and $\phi, \Omega$ are single-valued functions of the planar coordinates
$(x,y)$. 
 
We choose, as in Ref. \cite{zyg12}, 
\beq
&& \Omega(x,y) = \frac{\pi}{2} ( 1 + \tanh(\beta x)) \nonumber \\
&& \phi(x,y)=L \, B_{0} \, y
\label{S1.2}
\eeq
where $L, B_{0}, \beta$ are constants. 

In solving the Schroedinger
equation $ i \hbar \partial \psi / \partial t = H \, \psi $ it is useful to expand $\psi$ in the basis of the adiabatic eigenstates of $H_{ad}$. In that description we arrive at the set of coupled, Schroedinger-like,  equations for the multi-channel amplitudes that are minimally coupled to a pure, non-Abelian, gauge potential
$ {\bm A}({\bm R})  \equiv i \, U^{\dagger}({\bm R}) \, {\bm \nabla} \, U({\bm R})$, 
as well as the diagonal scalar potential matrix $H_{BO}$ whose entries are labeled by
 $\epsilon_{i}({\bm R}) $
and correspond to the Born-Oppenheimer energies of $H_{ad}$.
We now invoke the Born-Oppenheimer (BO) approximation and
project these coupled equations unto, open, ground-state BO amplitude $ F({\bm R})$. 
We get,
\begin{figure}[ht]
\centering
\includegraphics[width=0.7\linewidth]{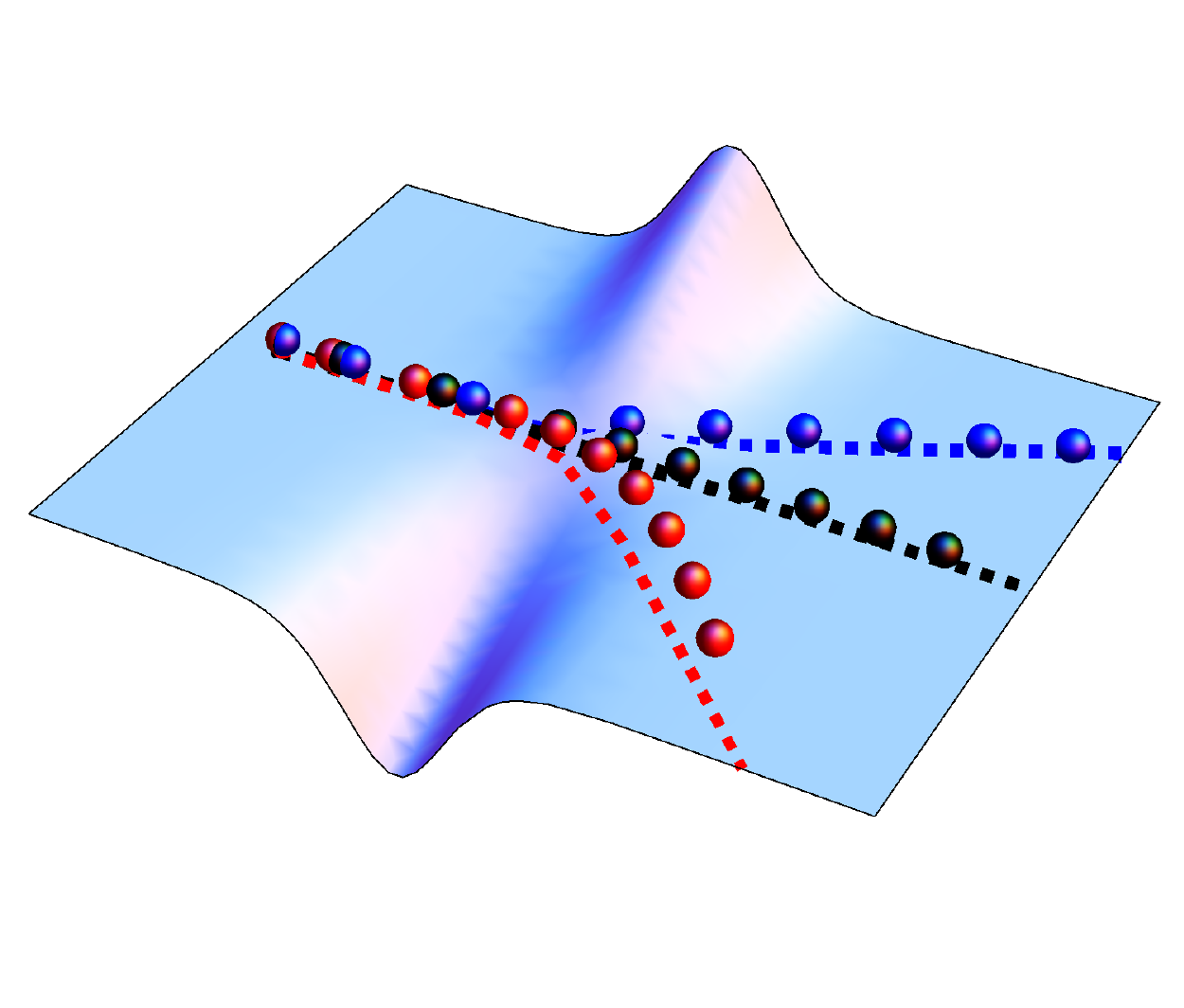}
\caption{\label{fig:fig1} (Color online) Classical trajectories (dotted lines) obtained from solutions to
Eq. (\ref{z0.9}) for various initial velocities. Colored spheres correspond to, time lapse, expectation values obtained  from solutions of the coupled channel quantum mechanical problem generated by Hamiltonian Eq.(\ref{z0.1a}). 
The black spheres pass undeflected with their initial velocities. }
\end{figure}
\beq
&& - \, \frac{\hbar^{2}}{2 m} \Bigl ( {\bm \nabla} - i {\bm A}_{P} \Bigr )^{2} F({\bm R}) + 
{\tilde V}_{BO}({\bm R}) F({\bm R}) = E F({\bm R}). \nonumber \\
&& \nonumber \\
&& 
{\tilde V}_{BO}({\bm R}) = \epsilon_{2}({\bm R}) +  b({\bm R}) \nonumber \\
&& b({\bm R}) \equiv  \frac{\hbar^{2}}{2 m} {\bm A}_{12} \cdot {\bm A}_{12} 
\label{z0.3}
\eeq 
where $b({\bm R})$ \cite{zyg86,zyg87a} is related, but not equivalent, to the adiabatic\cite{dal56} or Born-Huang correction\cite{BH}.
Typically the BO approximated is justified\cite{BH} if, for total (collision) energy $E$, the inequalities
$\epsilon_{1}({\bm R}) >> E > \epsilon_{2}({\bm R}) $, $ \Delta \epsilon \equiv \epsilon_{1} - \epsilon_{2} >> b({\bm R}) $ are
satisfied. 
${\bm A}_{P}$ is a vector potential and is obtained by projecting the non-Abelian gauge potential 
  so that ${\bm A}_{P} = Tr \, P {\bm A} P  \quad  P \equiv |g \rangle \langle g |$,
where $|g \rangle $ is the ground eigenstate of $H_{BO}$ with eigenvalue $\epsilon_{2}({\bm R})$, and
$ {\bm A}_{12}$ is the off-diagonal component of ${\bm A}$.
With this parameterization we obtain, for the ground adiabatic state,
an effective curvature 
\beq
&& {\bm H} =  \,  {\bm \nabla} \times \hbar {\bm A}_{P} = \nonumber \\
&&  {\bm {\hat k}}   \, \frac{\hbar}{4} \pi B_{0} L \beta \, {\rm sech}^{2}(\beta x) 
 \cos(\frac{\pi}{2} tanh(\beta x)).
\label{S1.3}
\eeq
Though ${\bm A}$ describes a pure non-Abelian gauge and so has vanishing
curvature, $ {\bm A}_{P}$ may have a non-trivial curvature, which for a non-degenerate ground state discussed here
has the Abelian form $ {\bm H} \equiv {\bm \nabla} \times {\bm A}_{P}$.

The classical limit for Eq. (\ref{z0.3}) corresponds to a situation in which the motion
of the atom is governed by\cite{zyg87a}
\beq
m \frac{d^{2} {\bm R}(t)}{dt^{2}} =  {\bm v} \times {\bm H}- {\bm \nabla}  {\tilde V}_{BO} 
\label{z0.7}
\eeq
where ${\bm R}(t)$ is the atom position coordinate. 

In addition to the conventional scalar gradient force 
$ - {\bm \nabla}  {\tilde V}_{BO} ({\bm R}) $, (sometimes called the Hellmann-Feynman force),
the atom experiences an effective velocity dependent Lorentz force. We argued\cite{zyg12,zyg12b} that such forces
could be exploited to construct effective ``magnetic lenses''  for neutral atoms or neutrons.
Here we underscore that observation by demonstrating that both the induced scalar
and vector forces could be used in conjunction to develop novel capabilities for the manipulation
and control of neutral particle beams. In particular, we describe a neutral particle velocity
sieve that exploits both the velocity dependent force arising from geometric magnetism as well
as the Hellmann-Feynman force.

It is well know that for charged particles a velocity selector can be realized by choosing a suitable
geometry in which a uniform magnetic field, ${\bm H}$, induces a Lorentz force that cancels 
the gradient force produced by an electric field, ${\bm E}$, 
so that  for a particle of charge $q$ and velocity $v$,  $ v q H = q E $. 
Pursuing this analogy
we construct a Hamiltonian of the form given in Eq. (\ref{z0.1a}) where
the entries of the BO eigenenergies are given by, 
\beq
H_{BO}=\sigma_{3}  \, (\epsilon({\bm R})+\Delta) -b({\bm R}) \nonumber \\
\epsilon({\bm R})=v_{0} \, y \, H(x) \quad H(x) \equiv  |{\bm H}|
\label{z0.8}
\eeq
and where $\Delta$ is chosen to be a sufficiently large energy gap so that the BO projection
approximation into the ground state is justified. In definition Eq. (\ref{z0.8}) we have included
a counter-term $b({\bm R})$  defined in Eq.(\ref{z0.3}). In the BO approximation  it's presence  cancels
the adiabatic correction generated by the off-diagonal components of ${\bm A}$.
With this choice, Eq. (\ref{z0.7}) becomes
\beq
m \frac{d^{2} x}{dt^{2}}= \frac{d y}{dt} H(x(t))+y(t) v_{0} \, \partial_{x}H (x(t)) \nonumber \\
m \frac{d^{2} y}{dt^{2}}= -\frac{d x}{dt} H(x(t)) + v_{0} H(x(t)).
\label{z0.9}
\eeq
In the asymptotic region $ x \rightarrow  -\infty$ all forces vanish and if we take the initial
condition $ {\dot y} =0, {\dot x}=v_{0}$,  
$ x(t) = v_{0} t + x_{0}, y(t)=0 $ is a solution to Eqs. (\ref{z0.9}). For other impact parameters and
velocities numerical solution of Eq. (\ref{z0.9}) predict trajectories in which the incoming
particle is scattered. This behavior is illustrated in Figure (\ref{fig:fig1}) where dashed lines
represent classical trajectories superimposed on the BO potential surface 
(ignoring the energy gap $\Delta $). The black line represents the solution in
which an atom with initial velocity $v_{0}$ propagates at constant velocity unimpeded.
The blue and red lines correspond to initial velocities slightly larger and smaller, respectively,
than $v_{0}$. Atoms with these velocities are clearly scattered by the effective Lorentz and
gradient forces. 

Below we demonstrate that this behavior is shared with quantum
evolution generated by Hamiltonian (\ref{z0.1a}). 
The atom is initially in its ground state in the asymptotic 
region and we propagate its amplitude using the time dependent method given in Refs.\cite{zyg12,zyg12b}. 
The packets are described by 
$$\psi \equiv \left ( \begin{array}{c} f \\ g \end{array} \right ) $$
whose components obey the coupled Schroedinger equations
\beq
&& i \hbar \, \frac{\partial f}{\partial t} = -\frac{\hbar^{2}}{2 m} {\bm \nabla}^{2} f + (V -b({\bm R})) \, f + V_{12} g \nonumber \\
&& i \hbar \, \frac{\partial g}{\partial t}= -\frac{\hbar^{2}}{2 m} {\bm \nabla}^{2} g + V^{*}_{12} f - (V + b({\bm R}) ) \, g
\label{S1.5}
\eeq
where 
\beq
&& \nonumber \\
&& V= (\Delta + \epsilon({\bm R}) ) \cos(\Omega(x)) \nonumber \\
&& V_{12}= \exp(-i \phi(y))  (\Delta+ \epsilon({\bm R}) ) \sin(\Omega(x)). \nonumber \\ 
\label{S1.6}
\eeq
In the remote past the packets 
have initial starting positions outside the interaction region with null impact parameters. In Figure (\ref{fig:fig1}) the black spheres represent propagation of packets with
initial velocity $v_{0}$. The centers of the packets track closely the classical trajectory. The blue
and red spheres have initial velocities slightly greater and smaller, respectively, than $v_{0}$.
The quantum simulation illustrated in that figure clearly demonstrates the proposed  velocity selection effect for
this system. Though Hamiltonian (\ref{z0.1a})  is a simple, time independent, two-state (or qubit)
system, it's experimental realization may pose challenges. Below we introduce another, more familiar, two-state system a neutral spin 1/2 system (eg. atom, neutron) subjected to a static external magnetic field. In the geometry discussed below we show how geometric phase induced, Aharonov-Bohm like, interferometry can be realized by it.

\subsection{Geometric phase, Aharonov-Bohm, interferometry}
\begin{figure}[ht]
\centering
\includegraphics[trim=0cm 7cm 0cm 7cm,clip=true,width=0.7\linewidth]{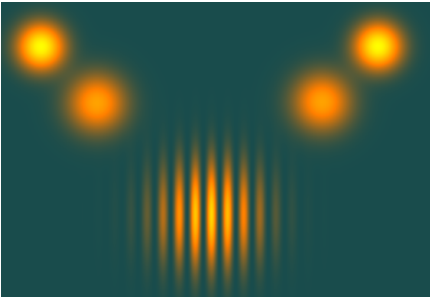}
\caption{\label{fig:fig2} (Color online) Time lapse illustration of a pair of coherent Gaussian wave packets initially in the ground state. At $\tau=0$ the pair of coherent packets are shown in the upper right and left sides
of the figure. At a later time, the packets coalesce thus creating the interference pattern. 
}
\end{figure}

Consider the external magnetic field 
\beq
 {\bm B}= B(\rho) \, {\bm {\hat \phi}} + B_{0} \, {\bm {\hat k} }
\label{z1.1a}
\eeq
where $\phi,\rho$ are the polar  and radial coordinates in a cylindrical
coordinate system, and $B_{0}$ is a constant, and we ignore motion in the $z$ directions as it can be factored from the planar motion. 
If $ B(\rho) = \lambda/\rho$ then Eq. (\ref{z1.1a}) describes
the field generated by a wire with current along the z-axis axis superimposed with that
of a homogeneous magnetic field $ B_{0} \, {\bm {\hat k} } $.
The Hamiltonian of a  neutral spin-1/2 atom is then given by
\beq
H= - \, \frac{\hbar^{2}}{2m} \, {\bm \nabla}^{2} + \mu_{B} \,  {\bm \sigma} \cdot {\bm B} 
\label{z1.1b}
\eeq
(alternatively, for a neutron $\mu_{B}$ is replaced by its magnetic moment).
We can 
re-express the internal Hamiltonian $H_{ad}=\mu_{B} {\bm \sigma} \cdot {\bm B}$  as
\beq
&& H_{ad}= \mu_{B} \left ( \begin{array}{cc} B_{0} & -i \, \exp(-i \phi) B(\rho) \\
                 i \, \exp(i \phi) B(\rho) & -B_{0} \end{array} \right ) = \nonumber \\
&&       U H_{BO} U^{\dag} 
		 \label{z1.2}
\eeq
where
\beq
H_{BO} = \mu_{B} \left ( \begin{array}{cc} \sqrt{B^{2}_{0}+ B^{2}(\rho)}  & 0 \\
                 0 & -\sqrt{B^{2}_{0}+ B^{2}(\rho)} \end{array} \right )
\label{S2.1}
\eeq	
and 
\beq
U =  \exp(-i \sigma_{3} \phi/2) \exp(i \sigma_{1} \Omega(\rho)/2) \exp(i \sigma_{3} \phi/2) 
\label{z1.1c}
\eeq
with $\phi$, $\rho$,  the azimuthal angle and radial distance, in a cylindrical coordinate system, respectively and $ \tan (\Omega) = B(\rho)/B_{0}$.
Thus, 
\beq
&& {\bm A}({\bm R})  \equiv i \, U^{\dagger}({\bm R}) \, {\bm \nabla} \, U({\bm R}) = 
{\hat  {\bm \phi} } \, A_{\phi} + {\hat  {\bm \rho} } \, A_{\rho} \nonumber \\
 && \nonumber \\
&& A_{\phi} = \frac{1}{2 \rho}  \left(
\begin{array}{cc}
 \cos \Omega (\rho )-1 &  i e^{-i \phi } \sin \Omega (\rho ) \\
 -i e^{i \phi } \sin \Omega (\rho ) &  1- \cos \Omega (\rho )  \\
\end{array}
\right)  \nonumber \\
&& A_{\rho} = -\frac{\Omega '(\rho )}{2}   \left(
\begin{array}{cc}
 0 &  e^{-i \phi }  \\
 e^{i \phi }  & 0 \\
\end{array}
\right)
\label{S2.2}
\eeq
and so,
\beq
&& {\bm A_{P}}= \nonumber   Tr \, P {\bm A} P =\frac{1-\cos\Omega(\rho)}{2 \rho}  \nonumber \\
&& {\tilde V}_{BO}= - \mu_{B} \, \sqrt{B_{0}^{2}+ B^{2}(\rho)} +  b(\rho) \nonumber \\
&& b(\rho)= \frac{\hbar^{2}}{2 m}  \Bigl ( \frac{\sin^{2}\Omega(\rho)/2}{ \rho^{2}} + \frac{\Omega'(\rho)}{4} \Bigr ).
\label{S2.3}
\eeq
The effective curvature for the ground adiabatic state is,
\beq
{\bm H} \equiv {\bm \nabla } \times \hbar \, {\bm A}_{P} = {\bm {\hat k}} \, \frac{\hbar}{2 \rho} 
\sin(\Omega(\rho)) \, \Omega'(\rho).
\label{S2.3a}
\eeq 
The above analysis is relevant in studies of the motion of cold atoms, that have a magnetic dipole moment, in the vicinity of current carrying  wire (or nanotube)\cite{zyg15}. 
Here we focus on a special case, that in which
$B(\rho) $ is a  constant $B_{\rho}$. For this case, according to Eq. (\ref{S2.3a}),
 ${\bm H}=0$, and because $V_{BO}(\rho)$ is also constant, and ignoring $b(\rho)$ 
(which is very small in the region $\rho \neq 0$
traversed by the packets), the atom does not
experience either a velocity dependent Lorentz, or scalar force. However, the induced vector potential
${\bm A}_{P}$ is not trivial and is given by
\beq
{\bm A}_{P}   = {\hat {\bm \phi}} \, \frac{\Phi}{2\pi \rho},  
\label{S2.4}
\eeq
where
\beq
 \Phi = \pi ( 1-\cos\Omega )= \pi ( 1- \frac{B_{0}}{\sqrt{B_{0}^{2}+ B_{\rho}^{2} }}) . 
\label{S2.4a}
\eeq
Packet propagation is again described as in Eq. (\ref{S1.5}) but now
\beq
&& V= \mu_{B} \, B_{0} \nonumber \\
&& V_{12}= -i\, \mu_{B} \, exp(-i \phi) B_{\rho}. \nonumber \\ 
\label{S2.5}
\eeq

\begin{figure}[ht]
\centering
\includegraphics[width=0.9\linewidth]{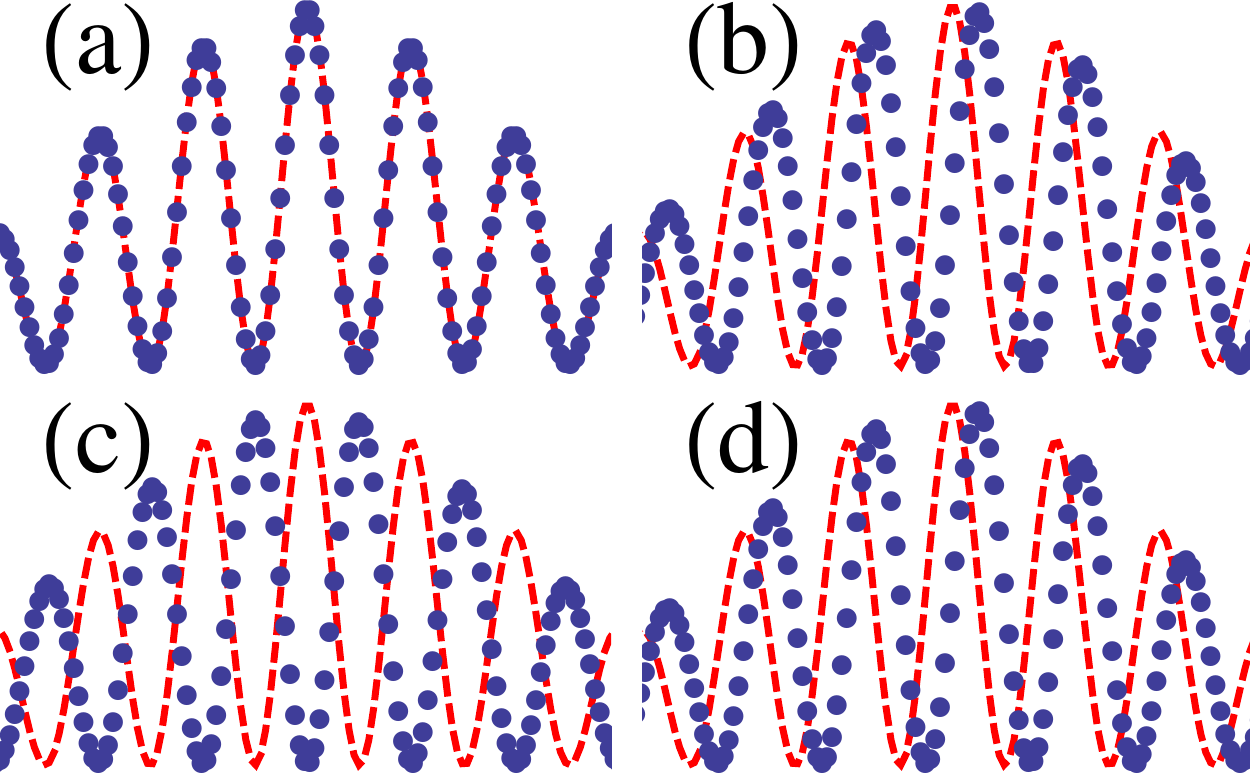}
\caption{\label{fig:fig3}(Color online) (a) Cross section of the interference pattern shown in Figure (\ref{fig:fig2}).
Red dashed line represents the analytic form Eq. (\ref{z1.9}). 
Blue dots represent results from numerical solution for the case $\Omega=0$. (b) Blue dots
represent interference cross section for the case $ \Omega=\pi/2$. (c) Same as above for $\Omega = \pi$,
(d) Same as above for $\Omega= 3\pi/2 $.}
\end{figure}

In order to demonstrate the proposed thesis we propagate two identical, coherent wave packets as shown in 
Figure \ref{fig:fig2}. 
The packets are displaced from the origin and are allowed to propagate, having
been given initial velocities that allow them to coalesce, at time $\tau_{c}$,  and form the interference pattern shown in
that figure. 
In our simulation we have first set $B_{\rho}=0$ so that the packets propagate freely.
In that case the horizontal line that passes through the center where
the two packets meet, the wave function has the analytic form (see Appendix A)
\beq
&& \frac{8 a^{2} k^{2} \exp(-\frac{ 2 a^{2} k^{2} \eta^{2}}{4 a^{4} k^{2} + \eta_{0}^{2}})}
{ 4 a^{4} k^{2} \pi + \pi \eta_{0}^{2} } \cos^{2}(k \eta) 
\label{z1.9}
\eeq
where $\eta$ is the horizontal coordinate, $k $ is a scaled wavenumber, $\eta_{0}$ are the initial displacements of
the packets from the origin and $a$ is the initial width of each packet. This function is plotted by the red line in Figure (\ref{fig:fig3}).
In panel (a) of that figure we superimpose the values, shown by the blue circles, obtained in our numerical simulation. 
We find excellent agreement with the analytic result Eq. (\ref{z1.9}) and validates the numerical procedure used in this
study. In panel (b) we plot the correspond interference pattern when $B_{0}, B_{\rho}$ has the value so that $\Omega=\pi/2$. We notice
a distinct shift in the calculated interference pattern from that given by Eq. (\ref{z1.9}).
 However, a fit with the replacement in Eq. (\ref{z1.9})
\beq
\cos^{2}(k \eta)  \rightarrow \cos^{2} (k \, \eta + \Phi/4)
\label{z1.10}
\eeq
provides an excellent approximation to the calculated data given by the simulation. 
 
\begin{figure}[ht]
\centering
\includegraphics[width=0.9\linewidth]{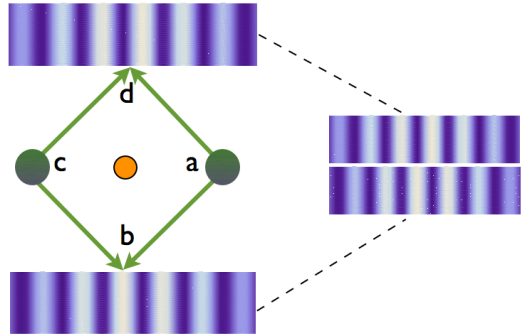}
\caption{\label{fig:fig4}(Color online) Fringe shifts produced by wave packet propagation of ground state solutions 
for the Hamiltonian given in  Eq. (\ref{z1.1b}) }
\end{figure}

We now consider
 the propagation of the two initial packets, at $t=0$, having their vertical velocities reversed
 so that they propagate into the upper half plane
and coalesce at a point that is the reflection of the coalescence point shown in Figure (\ref{fig:fig2}). 
The resulting interference pattern is illustrated in Figure ({\ref{fig:fig4}}).
At the point $d$ the pattern is again well described by Eq. (\ref{z1.9}) but with the replacement
\beq
\cos^{2}(k \eta)  \rightarrow \cos^{2} (k \, \eta - \Phi/4)
\label{z1.12}
\eeq
Because Hamiltonian (\ref{z1.1b}) is not invariant under reflection about the $\eta$ (horizontal) axis it is not
surprising that the interference pattern at point $d$ differs from that at point $b$. It might not be as obvious that
the difference is topological in nature.
According to the discussion above the locations on the $\eta$ axis where local minima occur
is given by
\beq
\eta_{m} = \frac{m \pi}{2 k} \pm  \frac{\Phi}{4 k}
\label{z1.13}
\eeq
where the $\pm$ sign identifies the points on the horizontal  lines passing through
the points  $b$ and $d$ respectively and $m$ is an integer. Therefore there is a displacement
\beq
\Delta \eta_{m} = \frac{\Phi}{2 k} 
\label{z1.14}
\eeq
shown by the right hand panel of Figure (\ref{fig:fig4}), in the location of
the relative minima between the upper and lower fringe patterns. It depends on the quantity
$\Phi$, which according to AB theory is given by
\beq
\int_{C} \, d {\bm r} \cdot {\bm A}_{AB}
\eeq
where $C$ is a contour that encircles the path $adcba$ in Figure(\ref{fig:fig4}) and 
$ {\bm A}_{AB}$ is given by Eq.(\ref{S2.4}). 
The connection with AB theory, and the topological nature of the fringe shift,
becomes evident when we shift the packet paths, so that a displaced closed circuit $a'd'c'b'a'$ no longer
includes the fictitious flux tube located at the origin. In that case our simulations show that the difference in the fringes
at the corresponding locations of $b',d'$ disappear, in harmony with the predictions of the gauge theory analysis.   

In the discussion above we presented the full quantum mechanical simulation of coupled equations (\ref{S1.5})
in order to derive the fringe patterns discussed above.  However, in order to gain additional insight it is useful to appeal to
semiclassical analysis of this system. Below we show how the results discussed above can be re-derived using
a semiclassical framework in the adiabatic representation of Eqs. (\ref{S1.5}).
\subsection{Semi-classical description of packet propagation in an Abelian gauge potential}

Consider the wave packet $\psi_{1} (\xi,\eta,\tau)$  defined by Eq. (\ref{S3.1})
and whose center, at $\tau=0$, is located
at point $a$ in Figure (\ref{fig:fig4}). We need to predict the packet that grows out of it and whose evolution is
determined by the coupled Schroedinger Eq.(\ref{S1.5}). Our calculation show that, under the adiabatic condition
$ \Delta/k^{2} >> 1 $, $\Delta \equiv \mu_{B} \, \sqrt{ B_{0}^{2}+B_{\rho}^{2} } $,  and in the adiabatic gauge, a good approximation for it at the time its center arrives at $b$ is 
\beq  U(a,b) \psi_{1}(b)  
\label{S4.1} \eeq
where $\psi_{1}$ is the free particle packet and the unitary operator $U(a,b)$ is given by
\beq
U(a,b) \approx \exp(i \int_{C_{1}}  d {\bm r}  \cdot {\bm A}_{P}). 
\label{S4.2}
\eeq
Here $C_{1} $ represent a path integral along segment $a-b$ that starts at $a$ and ends at $b$ and ${\bm A}_{P}$ is 
the gauge potential given by expression (\ref{S2.4}).  

Similarly, a packet initially centered at $c$ at $\tau=0$, translates
along path $c-b$ and arrives at $b$ at $\tau_{c}$. It can be expressed
\beq
 U(c,b) \psi_{2}(b)
 \label{S4.3}
 \eeq
where $\psi_{2}$ is defined in Eq. (\ref{S3.1}). The  coherent sum of these amplitudes at $\tau=\tau_{c}$ is then 
given by
the expression
\beq
&&  \psi(b)  =   U(a,b) \psi_{1} (b)+ U(c,b) \psi_{2}(b) = \nonumber \\
&& U(c,b) ( U^{-1}(c,b) U(a,b) \psi_{1}(b)  +   \psi_{2}(b) ) =\nonumber \\
&&  U(c,b) (  U(b,c) U(a,b) \psi_{1}(b)  + \psi_{2}(b) ) = \nonumber \\
&& U(c,b) (U(a,c)  \psi_{1}   +   \psi_{2}(b)) 
\label{S4.4}
\eeq
where we made use of the unitary property of $U$ and the relation $ U(a,c) = U(b,c) U(a,b) $.  
Therefore
\beq
| \psi(b)|^{2} =   | \psi_{2}(b) +U(a,c)  \psi_{1}(b) |^{2}
\label{S4.5}
\eeq

Evaluating 
\beq
U(a,c) =   \exp(i \int_{abc}  d {\bm r}  \cdot {\bm A}_{P} ) = \exp(-i \frac{\Phi}{2} ) 
\label{S4.6}
\eeq
and inserting this into Eq. (\ref{S4.4}) we obtain expression Eq. (\ref{z1.10}).

We obtain an analogous relation for the case where the momenta, along the $\xi$ direction, of the initial wave packets
at $a,c$ are reversed so that at time $\tau_{c}$ the packets meet at point $d$ in Figure 4. Following the steps outlined above we find
\beq 
| \psi'(d)|^{2}  =  |\psi'_{2}(d) + U'(a,c) \psi'_{1}(d) |^{2}
\label{S4.7}
\eeq
where $ \psi'_{i}$ are the corresponding free-particle packets whose $\xi$ momenta
are reversed, and
\beq
 U'(a,c) = \exp(i \int_{adc}  d {\bm r}  \cdot {\bm A}_{P} ) = U_{W} \, U(a,c) 
\label{S4.8} 
\eeq
where $U_{W}$ is a Wilson loop integral
\beq
U_{W} \equiv \exp(i \oint  d {\bm r}  \cdot {\bm A}_{P} ) =\exp(i \Phi)
\label{S4.9}
\eeq
 and the closed, counterclockwise, circuit encloses the origin.

According to  Eq. (\ref{S3.4})  
$ \psi'_{i}(\xi,\eta,\tau) = \psi_{i}(-\xi,\eta,\tau) $ and therefore, 
\beq
| \psi'(d)|^{2}  =  |\psi_{2}(b) + U_{W} U(a,c) \psi_{1}(b) |^{2}.
\label{S4.10}
\eeq
So the interference pattern $ | \psi'(d)|^{2}$ at the top panel in Figure (\ref{fig:fig4}), differs from that
at the lower panel by a, gauge invariant, phase determined by the Wilson loop $ U_{W} $, in harmony with
the results obtained by the, fully quantal, numerical simulation.

\begin{figure}[ht]
\centering
\includegraphics[width=0.9\linewidth]{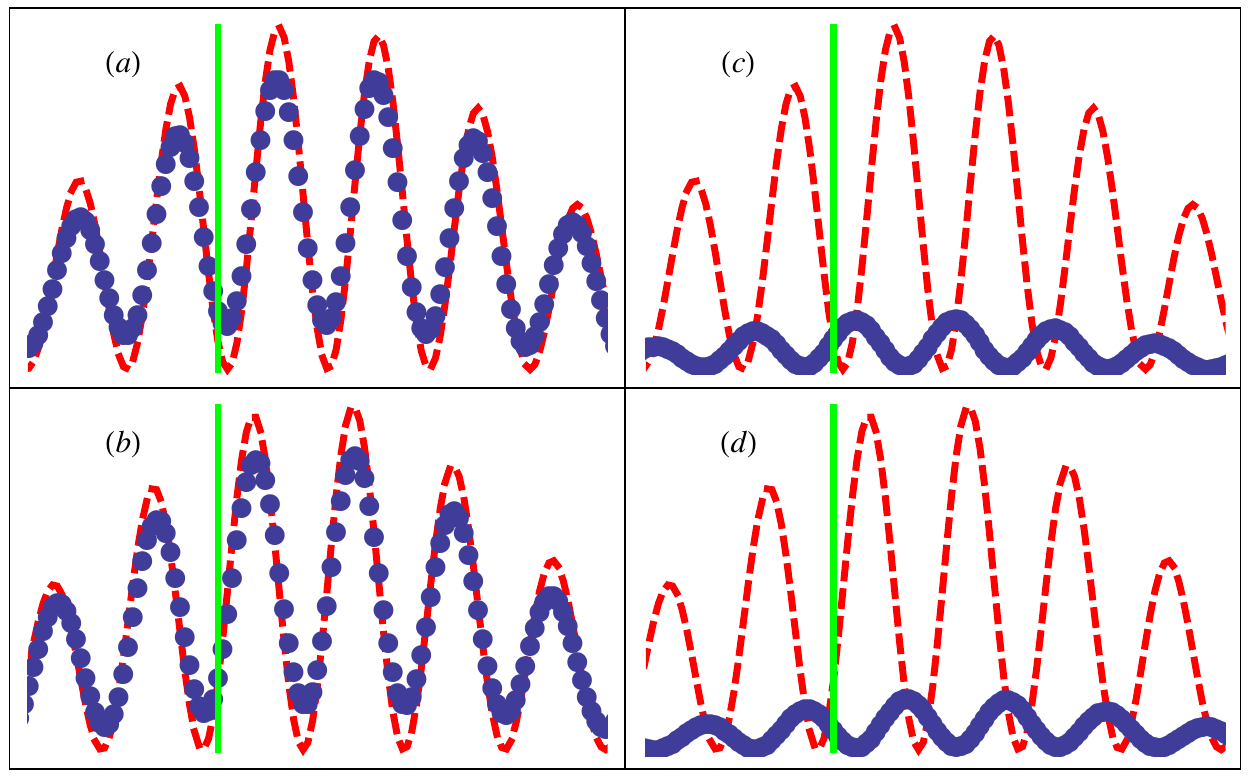}
\caption{\label{fig:figS1} (Color online) Interference fringes for energy defect  $\Delta/k^{2} =0.1$ and $\Omega=2\pi/3$. Panels (a),(b) are fringes (blue points) at locations b,d (bottom,top)  
in Figure (\ref{fig:fig4}) of text. Panels (c),(d) show fringes for (excited) state selected measurements at the latter locations. The green vertical line
is a reference line to aid the eye in comparing fringes between the top an bottom panels. The dashed red lines correspond to fringes predicted by the Abelian semiclassical theory}. \end{figure}

\begin{figure}[ht]
\centering
\includegraphics[width=0.9\linewidth]{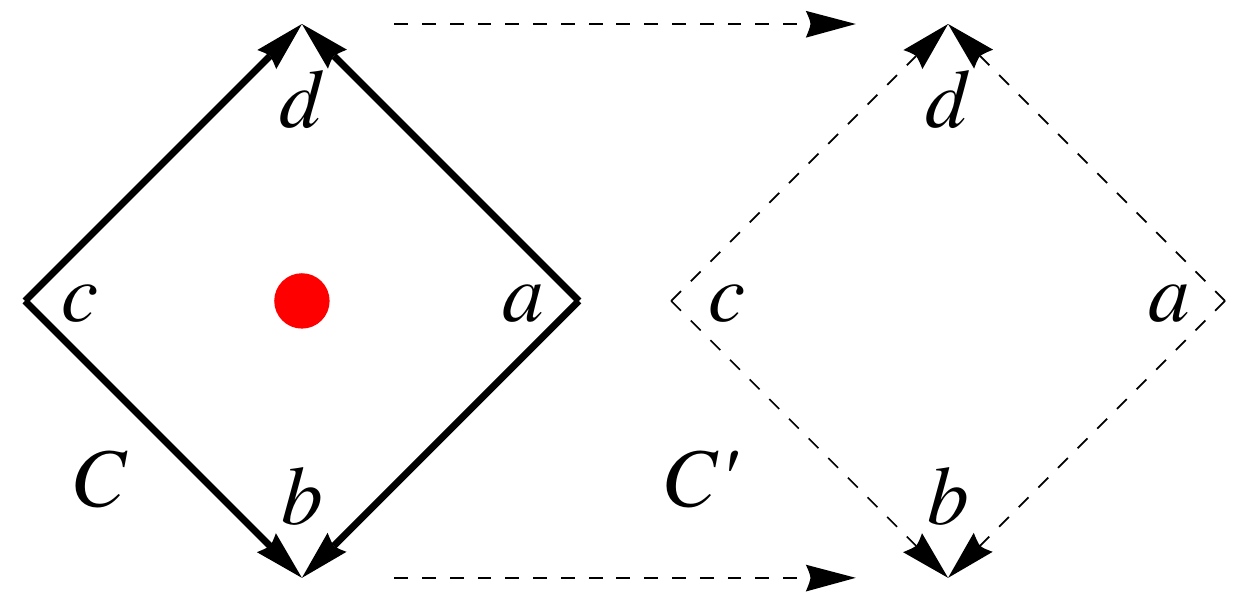}
\caption{\label{fig:figS2} (Color online) Circuit $ C $ which encloses fictitious flux tube (red disk), is translated to new circuit $ C'$ that does not enclose it. }
\end{figure}

\begin{figure}[ht]
\centering
\includegraphics[width=0.9\linewidth]{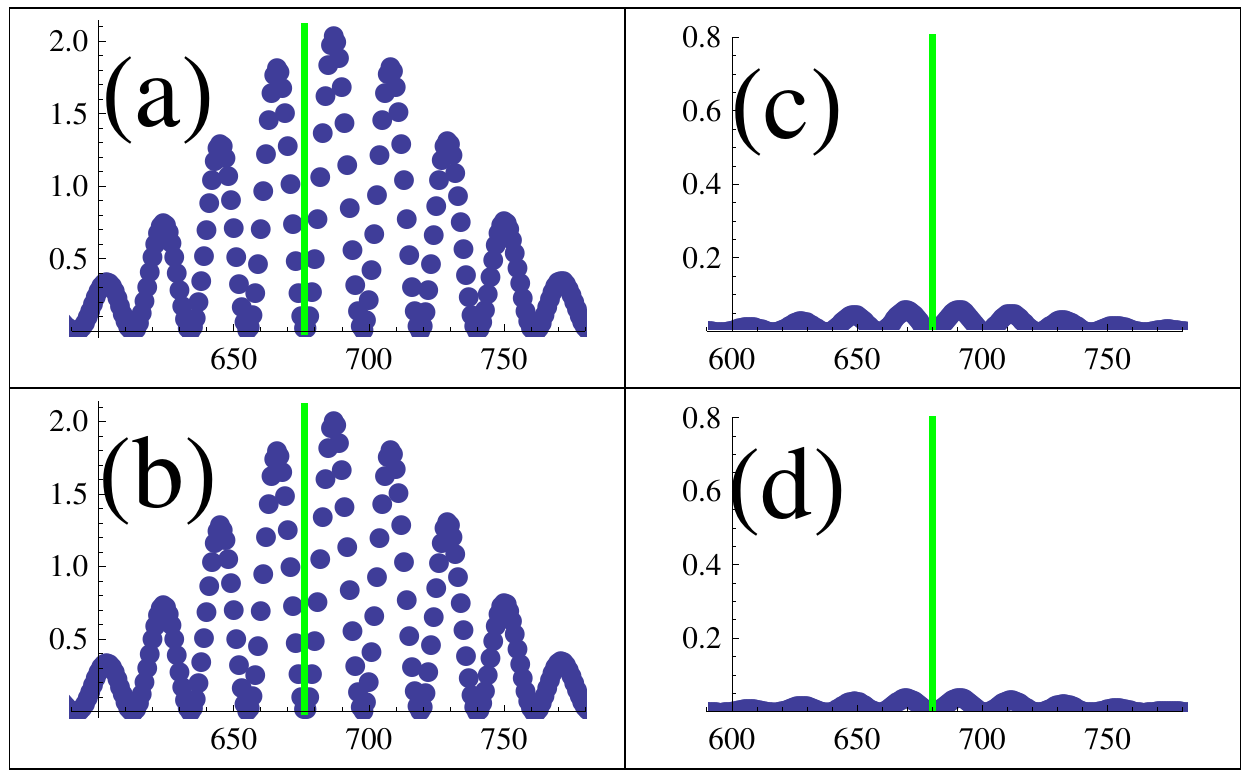}
\caption{\label{fig:figS3} (Color online) Interference fringes for energy defect  $\Delta/k^{2} =0.1$ and $\Omega=2\pi/3$
for loop diagram $C'$ in Figure (\ref{fig:figS2}). Panels (a),(b) show fringes (blue points) at locations b,d (bottom,top) in Figure (\ref{fig:figS2} )respectively. Panels (c),(d) show fringes for (excited) state selected measurements at points b,d (bottom,top) in Figure (\ref{fig:figS2}) respectively. The green vertical line is a reference line to aid the eye in comparing fringes between the top an bottom panels. } 
\end{figure}

\subsection{Semi-classical description of packet propagation in a, pure, non-Abelian gauge potential}
If the collision energy, $ \frac{\hbar^{2} k^{2}}{2 m} $, is much larger than the energy defect $ 2 \Delta$, between
the Zeeman split spin states, non-adiabatic transitions between those states can occur. Thus if the initial
localized wavepacket, say  at point $a$ in Figure (\ref{fig:fig4}),  describes a particle in the ground Zeeman level,
it will not necessarily stay in that level as the packet evolves in time.

In the discussion above we considered the adiabatic limit in which the ratio of energy defect to collision energy is
large i.e. $ \Delta/k^{2} >> 1 $. In that limit spin flipping transitions between ground and excited Zeeman levels are
suppressed. We showed that, in this limit, the single channel (or Abelian) Schroedinger equation (\ref{z0.3}) with gauge potential Eq. (\ref{S2.4}) and $ V_{BO}=0$ accurately predicts  wavepacket dynamics and topological AB features. 
As the collision energy is cranked up so that  $ 2\Delta/k^{2}  \leq 1 $  we anticipate that the
single channel (Abelian) description breaks down and non-Abelian features arise.

In Figure (\ref{fig:figS1}), panels (a), (b)  we plot the interference patterns, corresponding to the top and bottom regions shown in Figure (\ref{fig:fig4}) for the collision energy corresponding to $\Delta/k^{2} =0.1$. Though the collision energy is sufficient to cause Zeeman level transitions, our results suggest that
many of the Abelian features persist. First we note that there is a phase shift between the interference patterns, for the top and bottom regions respectively, that is nearly, but not exactly, predicted by the Abelian AB theory (which are shown in red in that figure). Panels (c),(d) show the fringes, for the top and bottom regions respectively for state-dependent  probabilities,
in this case for excitation into the upper Zeeman level.  A distinct phase shift in the fringe patterns
for excitation is also seen, though its structure is not predicted by the Abelian theory.

In order to investigate whether these features are topological we 
translate the loop $C$, shown in Figure (\ref{fig:figS2}), into the loop $C'$  and repeat the calculations
described above. Because loop $C'$ does not enclose the fictitious flux tube (shown by the red disk) classical
AB theory suggests that the difference in fringe shifts, evident in Figure (\ref{fig:figS1}), 
is null.
Indeed, this is the case. In panels (a),(b), of Figure (\ref{fig:figS3}), the probability interference are
shown. The patterns for top and bottom regions (points d,b respectively) line up and no fringe shifts are evident.
Panels (b),(d) of that figure show the corresponding interference patterns for the excited Zeeman level
probabilities. Though fringe differences are negligible, we note a strong suppression of the latter 
(when compared to that for loop $C$ shown in panels (b),(d) in Figure (\ref{fig:figS1}) ). The suppression of
excitation for loop $C'$ is clearly a non-Abelian (or multichannel) feature.

Finally, we consider the extreme non-adiabatic regime. In it the ratio $ \Delta/k^{2} \rightarrow 0$ 
and we can again employ semi-classical methods\cite{zyg15} to predict the fringe patterns that are
generated by our fully quantal simulations and which are shown below. In Figure (\ref{fig:figS4}) we repeat the
calculations for propagation along loop $C$ in Figure (\ref{fig:figS2}) for the values
$ \Delta/k^{2} =0.007 $ and $ \Omega=2\pi/3$. Unlike the case for the adiabatic and near adiabatic 
regimes, in which topological fringe shifts arise, panels (a), (b) of Figure (\ref{fig:figS4}) 
clearly demonstrate absence of the topological fringe shift. This behavior can be explained 
using semiclassical methods.  For, in that description\cite{zyg15}
the total probability amplitude 
\begin{figure}[ht]
\centering
\includegraphics[width=0.9\linewidth]{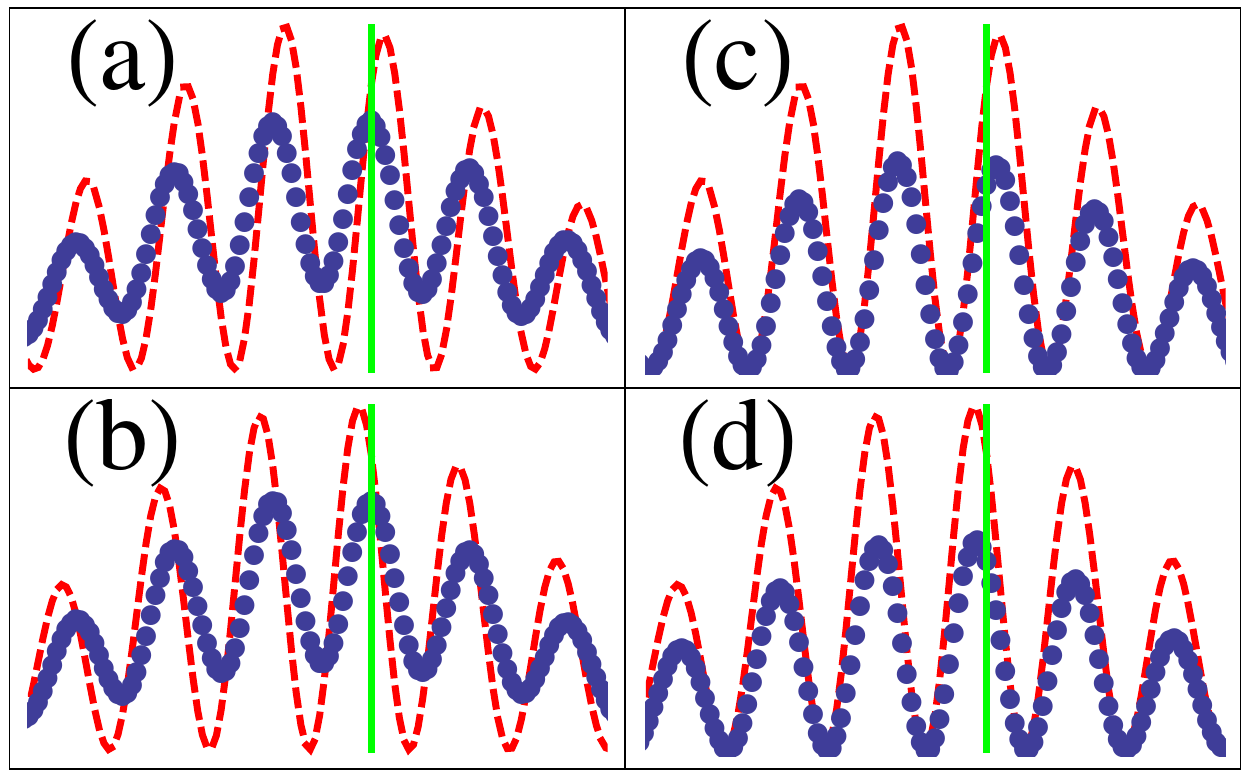}
\caption{\label{fig:figS4} (Color online) Interference fringes for energy defect  $\Delta/k^{2} =0.007$ and $\Omega=2\pi/3$. Panels (a,b) fringes (blue points) at locations b,d (bottom,top)  in Figure (\ref{fig:fig4} ) respectively. Panels (c,d) fringes for (ground) state selected measurements at points b,d (bottom,top)  in Figure (\ref{fig:fig4} ) respectively. The green vertical line
is a reference line to aid the eye in comparing fringes between the top an bottom panels. The dashed red lines
correspond to fringes predicted by the semiclassical Abelian theory}. 
\end{figure}
at point (b) in (\ref{fig:fig4}), that grows out of wave packet $\psi_{1}(a)$ is approximated by the expression 
\beq
 P \exp(i \int_{ab} d {\bm r} \cdot {\bm A})
\psi_{1}(b)
\label{S4.10b}  
\eeq
where $P$ is a path-ordered integral, or Wilson line, along segment $a-b$, and $ {\bm A}$ is the non-Abelian, pure, gauge potential (\ref{S2.2}).
Repeating the argument outlined above for the fringe shift in the adiabatic limit, we now find that the shift
depends on the Wilson loop integral
\beq
U_{W} \equiv P \exp(-i \oint  d {\bm r}  \cdot {\bm A} )  
\label{S4.11}
\eeq 
where ${\bm A}$ is given by  Eq.(\ref{S2.2}). Because $ {\bm A}$ describes a pure gauge we find that $U_{W}=1 $ (see Appendix C) 
and so, unlike the case in the adiabatic regime,
a fringe shift between the interference patterns at $b,d$ in (\ref{fig:fig4}) does not manifest. Interestingly, this is no longer true if we perform state dependent
 measurements at locations $b$, $d$ in that figure. In panels $(c),(d)$ of Figure (\ref{fig:figS4}) we plot the calculated interference patterns at those points
 for a state selective measurement (in this case, the ground state). Those fringe shifts are, again, accurately predicted by the Abelian theory as illustrated
by the red lines in those panels. Furthermore, this shift is also topological, in that the shifts vanish for loops that do not enclose the fictitious flux tube.
A detailed discussion of the origin and implications of this observation will be presented elsewhere\cite{zyg15}.
\subsection{AB inteferometry for a single loop}
\begin{figure}[ht]
\centering
\includegraphics[width=0.4 \linewidth]{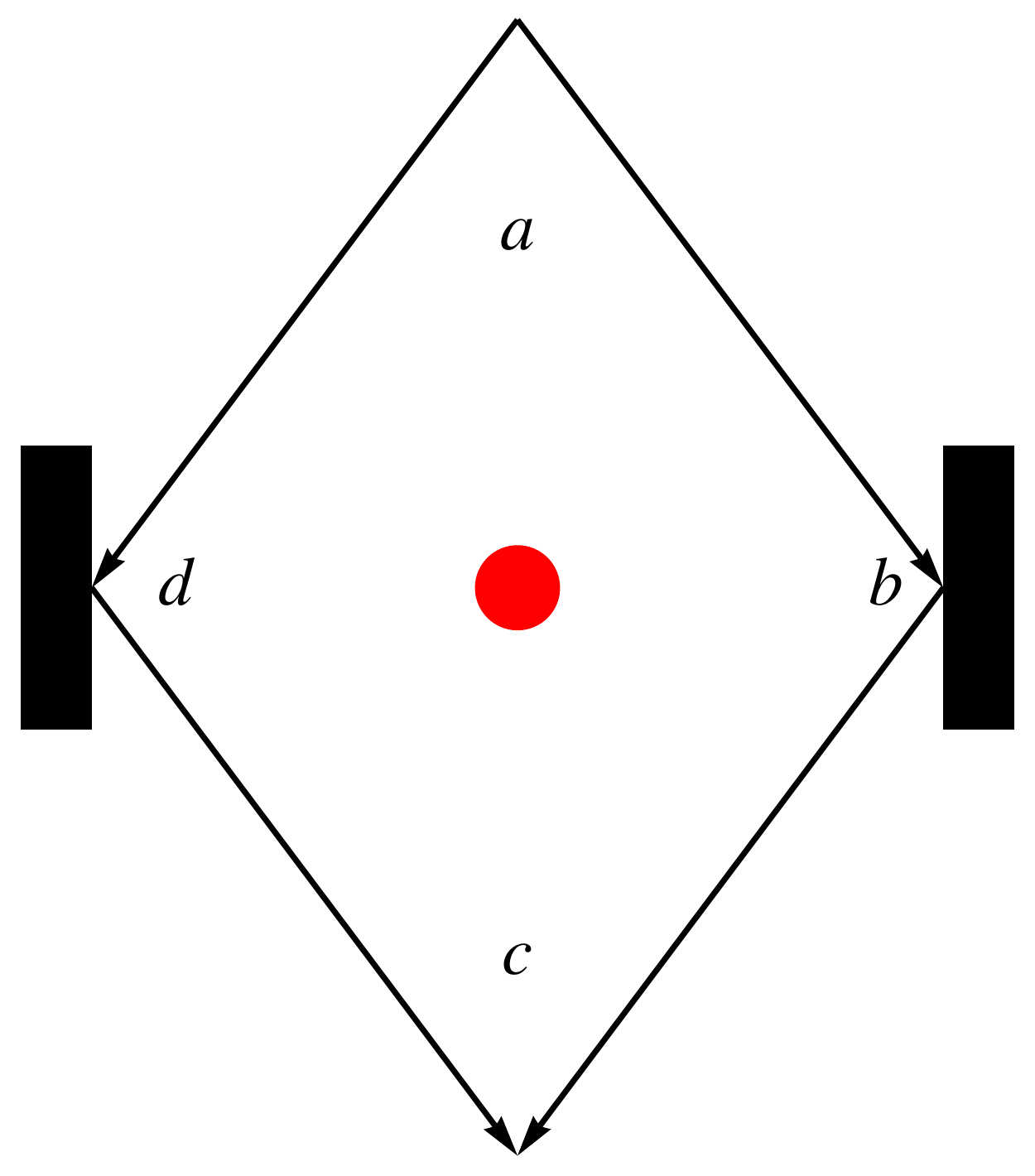}
\caption{\label{fig:figS5} (Color online) Standard, single loop, AB inteferometry setup. Two coherent packets at origin $a$ split and propagate to the mirrors at points $b$, $d$. The mirrors deflect the packets so they recombine at point $c$ where the measurements are made.}. 
\end{figure}

Our discussions addressed AB inteferometry for setups in which interference patterns
are compared following two independent open-loop measurements\cite{aharonov06}. In the classical single loop
AB setup, an interference pattern is observed at a single screen (point $c$) as shown in Figure
(\ref{fig:figS5}). In it, a wavepacket coherently splits at the origin, point $a$, and 
propagates toward the mirrors at points $b$,$d$ respectively, the packets are deflected and 
allowed to recombine at point $c$ where
a measurement is taken. In the calculations described above, and in the adiabatic limit
$ \Delta/k^{2} >> 1 $, we found that ($i$) the wave
packets propagate as a free particle, ($ii$) in the journey along curve $C$ the
packets acquire, in addition to the standard dynamical the phase factor, the phase
\beq
 \exp(i \int_{C} d {\bm r} \cdot  {\bm A}_{P}). 
 \label{S5.1}
 \eeq
Using Eq. (\ref{S5.1}) for the paths $a-b-c$ and $a-d-c$ we find that measurements
at the screen will be a function of the path integral, along the closed loop, 
\beq
\exp(i \oint d{\bm r} \cdot {\bm A}_{P}).
\label{S5.2}
\eeq 
If that loop encloses the fictitious flux tube then its value is $\exp(i \Phi)$, otherwise it has unit value.   Therefore this
setup 
exhibits topological properties consistent with AB theory.
In the extreme non-adiabatic limit $ \Delta/k^{2} << 1 $ our calculations again demonstrate the validity of
properties $(i),(ii)$, with the exception that  ``free particle'' evolution is
that of a two-component wavefunction and the phase factor multiplying it  is a multi-channel unitary matrix
Eq. (\ref{S4.10b}). Therefore, repeating the analysis given above we find 
measurements at screen $c$  are now proportional to the loop integral 
\beq
P \exp(i \oint d{\bm r} \cdot {\bm A}) = 1.
\label{S5.3}
\eeq
That is, regardless of the loop geometry, topological fringe patterns do not arise and there is no topological AB shift.
In summary, our calculations demonstrate that, in the adiabatic limit, solutions generated
by the time evolution operator Eq. (\ref{z1.1b}) with the external field configuration Eq.(\ref{z1.1a}) 
reproduces the standard AB fringe shift. In the limit $\Delta/k \rightarrow 0$, non-adiabatic
transitions conspire to wash out topological AB fringes. This conclusion is consistent with that given in Ref. \cite{march92}.
Neverthless, we find here that if spin-state dependent measurements are made during traversal of the 
circuit, topological shifts persist. This counter-intuitive
observation will be discussed in more detail elsewhere\cite{zyg15}.

\section{AB scattering}
In the previous section we presented an outline for eliciting AB-like topological effects, for neutral spin 1/2 particles, in an interferometry setup. In standard treatments e.g. \cite{Berry80,Au84},  the AB effect is discussed in the context of a scattering scenario and
 so it
is instructive to analyze the dynamics generated by Hamiltonian Eq.(\ref{z1.1b}) in this framework.
In Ref.\cite{march92}  a Born-Oppenheimer approximation was used to obtain an equation for the ground state
amplitude (in a field theoretic analog of this system), and in which, the vector potential ${\bm A}_{AB}$ 
 emerges.  Standard time independent methods were applied to demonstrate AB -like  scattering. 
 A BO approximation was also used in Ref.\cite{shen96} in order to suggest that AB -like scattering arises in systems described by Hamiltonian (\ref{z1.1b}).

Here we apply time dependent wave propagation in order to demonstrate AB scattering. However, there are two obstacles that prevent us from simulating  pure AB scattering in the Frauenhofer region. The latter requires
propagation of a wave packet from the distant remote past, to the distant future. This is numerically untenable. In addition, the wave packet traverses a region near the origin where singular couplings between the ground an excited adiabatic states
arise\cite{zyg15}. We by-pass these difficulties by placing an impenetrable hard-cylinder  surrounding the origin
and introducing a diagonal  counter term 
$b_{ii}({\bm R})  =  \sum_{i \neq j} {\bm A}_{ij}\ \cdot  {\bm A}_{ji}   $, as shown in Eq. (\ref{S1.5}).

\begin{figure}[ht]
\centering
\includegraphics[width=0.6 \linewidth]{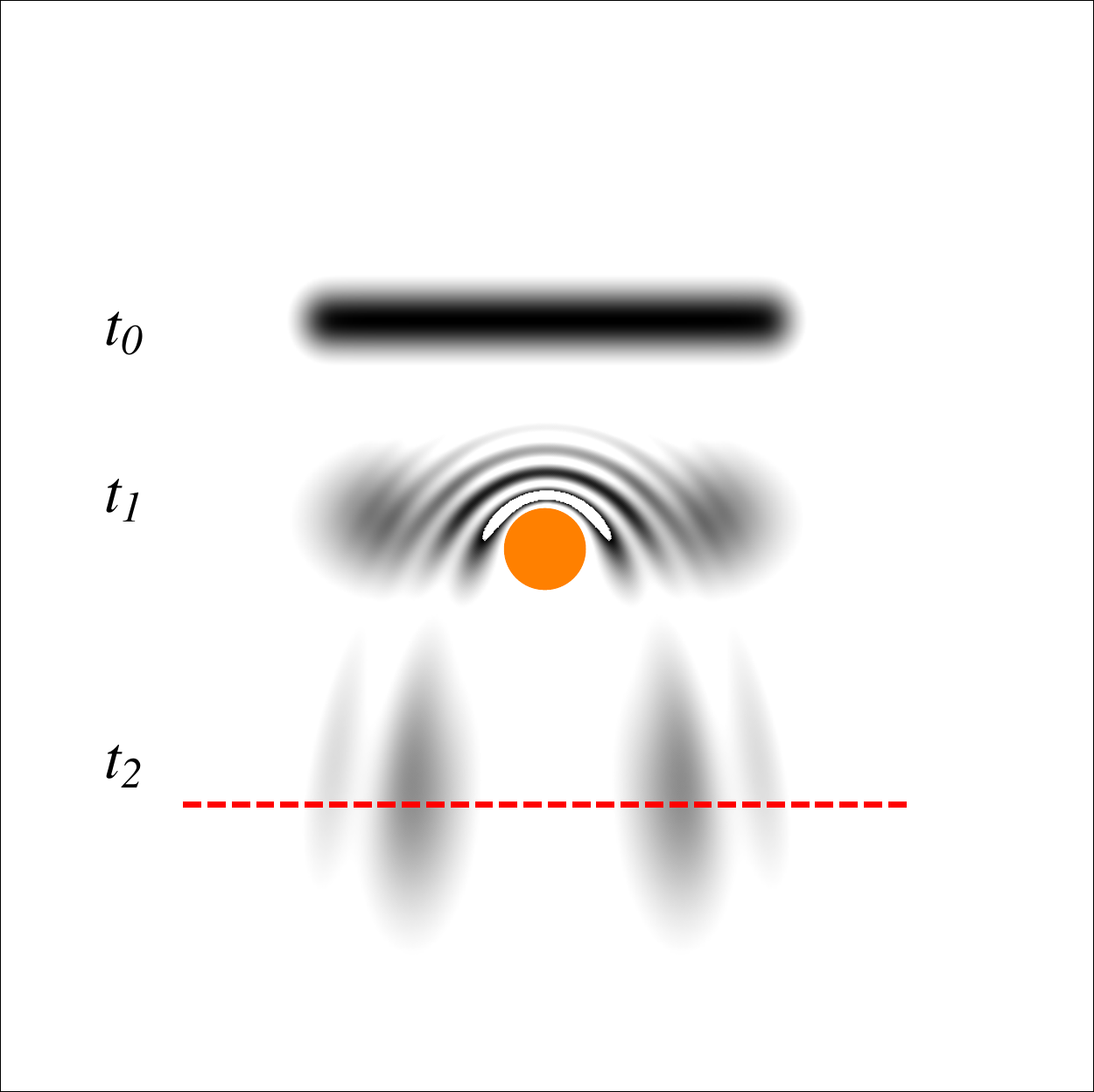}
\caption{\label{fig:fig10} (Color online) Time series plot of the probability density for an initial wavepacket at $t_{0}$ that propagates toward the impenetrable cylinder
(orange disk) and is scatterd by it. The red dashed line represents a detection screen.} 
\end{figure}

At initial time $t_{0}$ we introduce a finite slab-like wave packet  shown by the illustration
in Figure \ref{fig:fig10}. It proceeds with a mean initial velocity toward the impenetrable cylinder represented in
that illustration by the orange colored disk. The wave packet is propagated numerically solving
coupled equations (\ref{S1.5}). The packet mean momentum is chosen 
so that the inequality $ \Delta/k^{2} >>1$
is satisfied. At a later time $t_{1}$ Figure \ref{fig:fig10} illustrates how the packet diffracts around the
cylinder. Subsequently, at $t_{2}$, the forward scattering probability density reaches a detector shown
by the red dashed line in that figure.

\begin{figure}[ht]
\centering
\includegraphics[width=0.9 \linewidth]{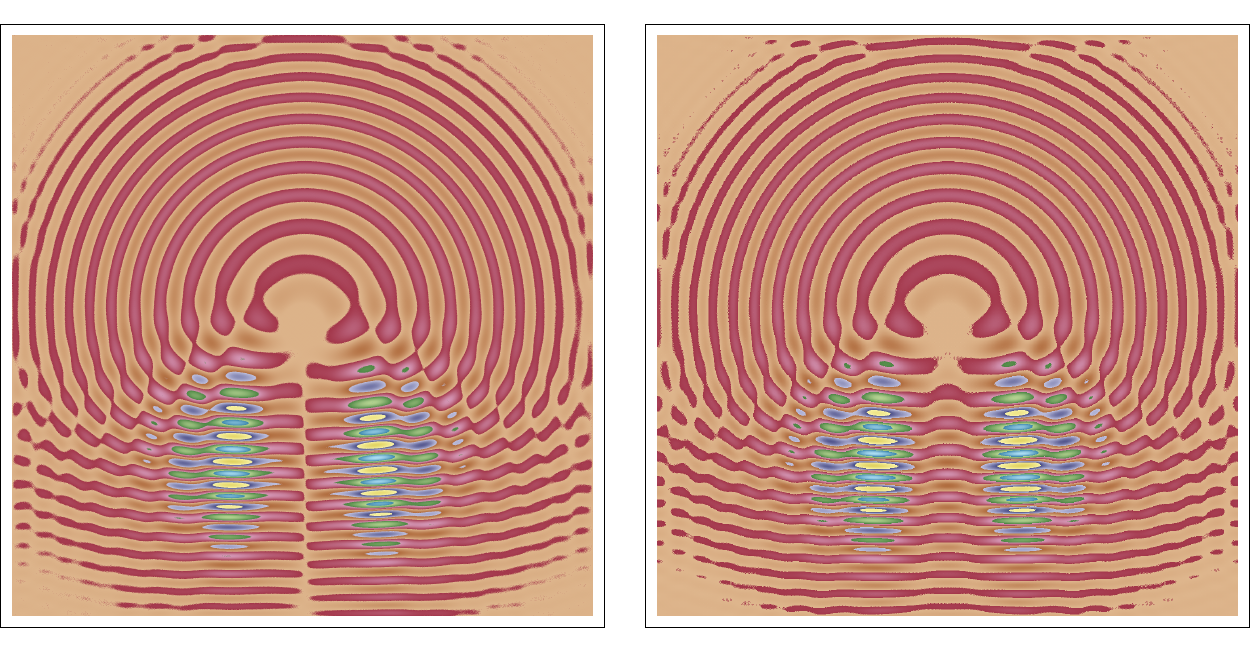}
\caption{\label{fig:fig11} (Color online) Plots of the imaginary part of the ground state amplitude at time $t_{2}$
shown in Figure (\ref{fig:fig10}). Right panel corresponds to case $\Phi=0$, scattering by an impenetrable cylinder. The left
panel corresponds to the case $\Phi=\pi$, maximal AB scattering by an impenetrable cylinder.} 
\end{figure}

\begin{figure}[ht]
\centering
\includegraphics[width=0.9 \linewidth]{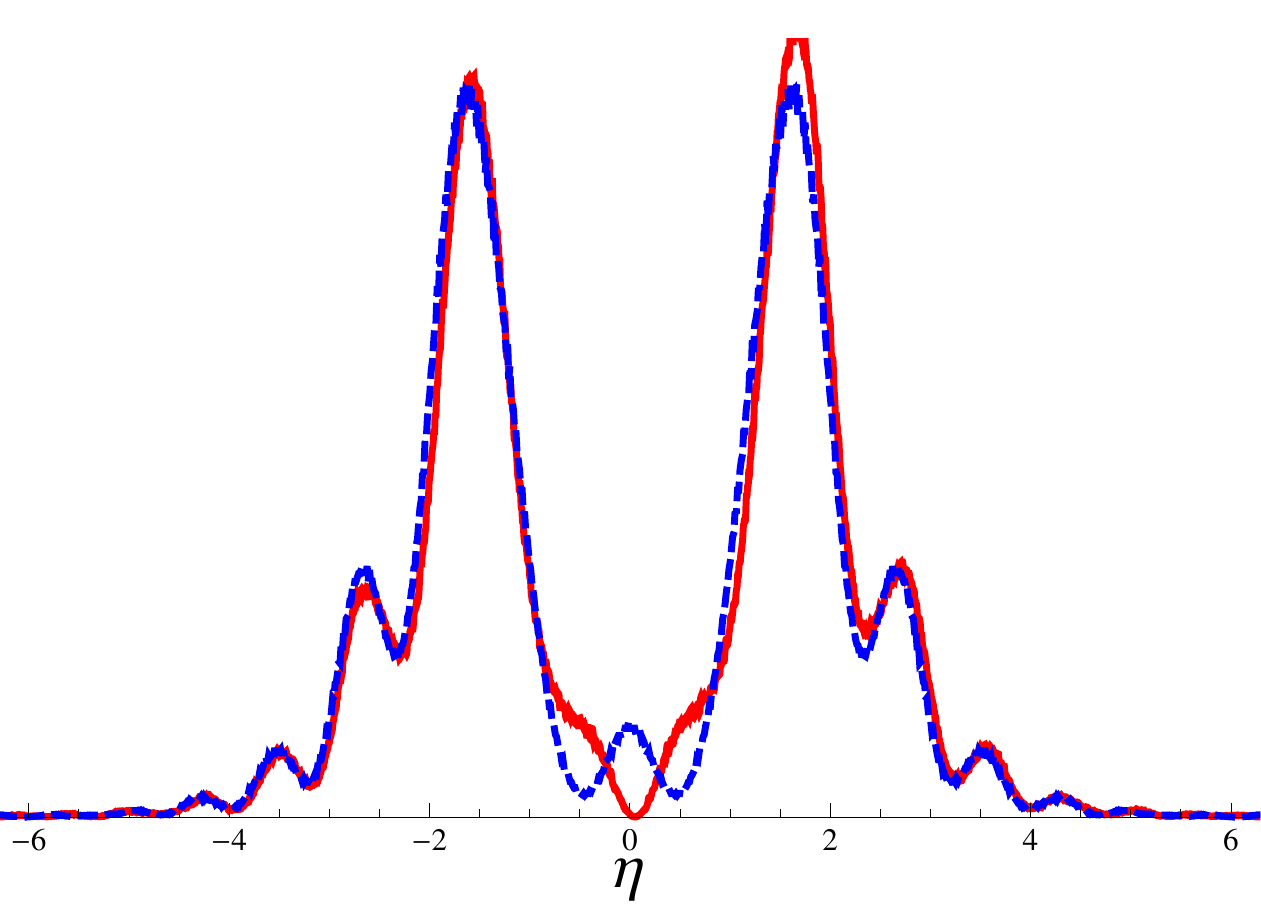}
\caption{\label{fig:fig12} (Color online) 
Dashed blue line is a plot of the time dependent 
probability density, at $t_{2}$ and for $\Phi=0$, along the $ \eta $ axis corresponding to the red dashed line shown in Figure (\ref{fig:fig10}).
The red solid line corresponds to the case $\Phi=\pi$, maximal AB scattering by an impenetrable cylinder.}
\end{figure}

In Figure \ref{fig:fig11} we present a high resolution plot for the imaginary part of the wave packet amplitude, at time $t_{2}$, for the two cases where $\Omega=1/2$ (left panel) and $\Omega=0$ (right panel) respectively. With $\Omega =0$ Eqs. (\ref{S1.5}) decouple,and the ground state
amplitude satisfies the free particle Schroedinger equation in the presence of a impenetrable cylinder centered at the origin. The right hand panel of Figure \ref{fig:fig11}  illustrates both the reflected and transmitted components, including interference between the two, of the packet as it is scattered by the cylinder. In the left hand panel we plot the same amplitude except we set the parameter $\Omega=1/2$ in Eq. (\ref{S1.6}). It corresponds to the case where $\Phi=\pi$ in Eq. (\ref{S2.4}).
The plot is almost identical to the one obtained for $\Omega=0$, except for the striking phase
dislocation\cite{Berry80}, running along a line bisecting the cylinder in the forward scattering direction. This phenomenon is well
know in pure AB scattering\cite{Berry80} and, for the case $\Phi=\pi$, is a manifestation of a nodal line for the amplitude along the phase dislocation line. Berry et al.\cite{Berry80} argued that phase dislocation in pure AB scattering, though not a
physical observable,  is a topological invariant. 
In Figure (\ref{fig:fig12}) we plot the probability amplitude at the observation panel at time $t_{2}$, shown by the line in Fig. (\ref{fig:fig10}), for both
cases $\Omega=0$, the blue dashed line, and $ \Omega=1/2$ by the solid red line. Both probability distributions are similar except
along the line near $\eta=0$ in that figure. Both cases exhibit the ``shadow'' cast in the forward
direction by the cylinder, the $\Omega=0$ plot show a small enhancement directly "downwind", at $\eta=0$, 
from the cylinder. It corresponds to the 1D analog
of the celebrated Poisson spot\cite{poisson98,Poisson09} that occurs when a wavefront diffracts about a circular obstacle. The red line in Fig. (\ref{fig:fig12}), which corresponds to the case where $\Phi=\pi$ in AB scattering and which shows a strong suppression of this spot. 

Finally, we compare the results shown in Fig. (\ref{fig:fig12}) with those obtained in a time independent description.
In appendix B, we construct the wave functions used in a standard scattering theory scenario. It obeys  boundary
conditions that correspond to an incoming packet in the asymptotic region in the remote past that subsequently scatters off the cylinder. Typically this wave function is used to derive the scattering amplitude in the Frauenhofer region, but here we use
it to find the, time independent, probability amplitude at the observation screen shown in Fig. (\ref{fig:fig10}). In Figure (\ref{fig:fig12}) the dashed blue
line represents the case $\Phi=0$ (i.e. no AB flux tube), and  the red line by the $ \Phi =\pi$ case.
The latter corresponds to maximal AB scattering. As in Fig (\ref{fig:fig12}),  differences in the probability distributions are most
evident near the $\eta \approx 0$ line. Because the incident amplitude extends from $ -\infty < \eta < \infty $,  Fresnel type
interference patters at larger $\eta$ manifest here and which are not present in Fig. (\ref{fig:fig12}). However, there is qualitative agreement
between the two descriptions near the forward direction, in that a strong suppression of the Poisson spot is evident
when the AB flux tube has the value $\Phi=\pi$. 

In summary we have demonstrated that solutions of Eq. (\ref{S1.5}) do exhibit the features that arise in standard AB scattering
for the collision energy range $ \Delta/k^{2} >>1$. 
 
\begin{figure}[ht]
\centering
\includegraphics[width=0.9 \linewidth]{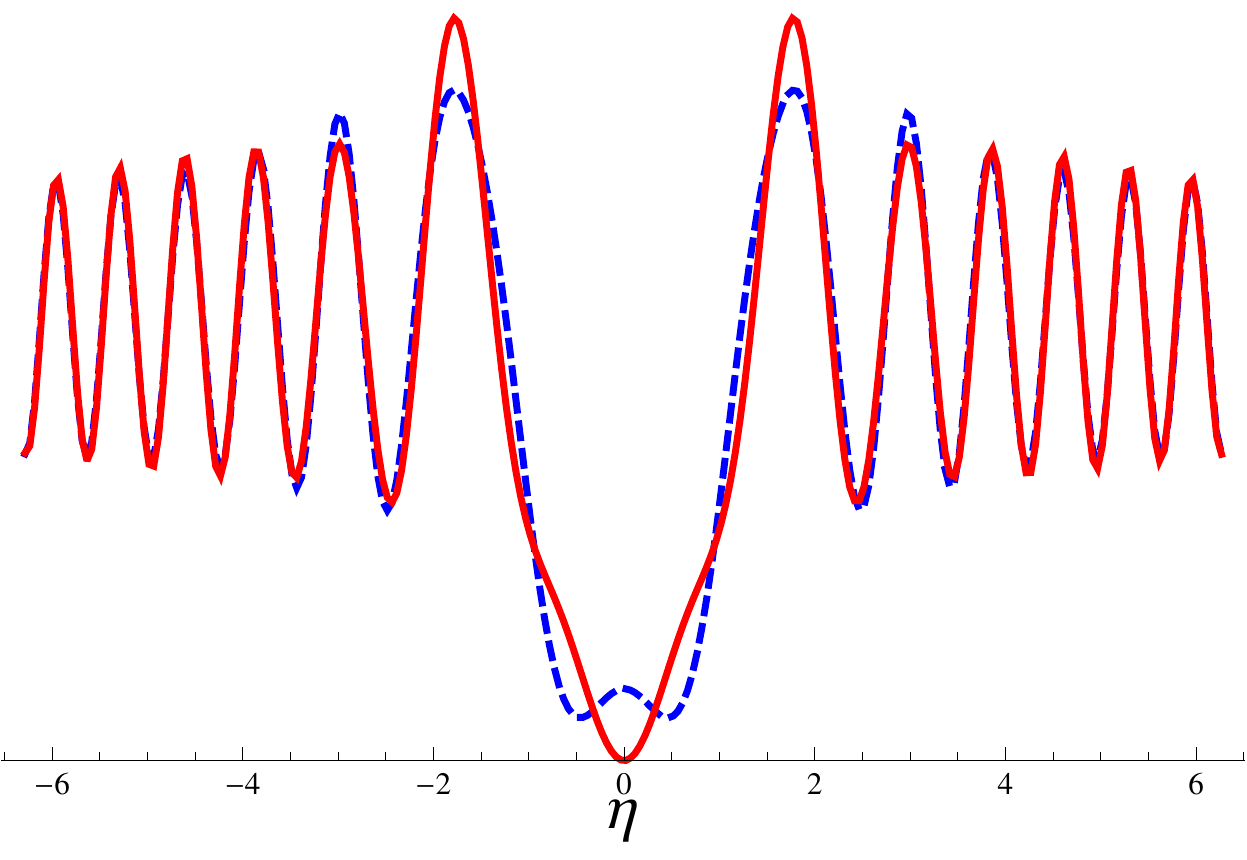}
\caption{\label{fig:fig13} (Color online) 
Dashed blue line is a plot of $ |\psi |^{2} $ along the $ \eta $ axis, corresponding to the detection screen shown in  figure (\ref{fig:fig10}), for solution to Eq. (\ref{1.0}) with $ \Phi=0$. It corresponds to time independent scattering from an impenetrable cylinder at the origin.  The red solid line corresponds to the case $\Phi=\pi$,  maximal AB scattering by an impenetrable cylinder.}
\end{figure}

\section{Relation to the molecular AB effect}
A gauge potential equivalent to Eq. (\ref{S2.4}) (for $\Phi=\pi $) 
was introduced by Mead and Truhlar \cite{mead76} 
in order to describe molecular dynamics near a conical intersection. In this section we explore the relationship between that
phenomenon, often called the molecular Aharonov-Bohm (or MAB) effect, and  AB-like scattering described in the previous section. It has long been taken for granted, in the molecular
physics community, that degeneracy in the form of a conical intersections is an essential requirement for topological
effects induced by a vector potential to arise.  However, no degeneracy in the physical parameter space
${\bm R}$ is evident in the BO Hamiltonian defined in Eq. (\ref{S2.1})  as $B_{0},B_{\rho}$ are taken to be constant.
This dichotomy presents a certain amount of cognitive dissonance and so deserves closer examination.

Consider a tri-atomic system that possesses a conical intersection at the origin of a parameter space that is spanned by a set
of nuclear internal coordinates  $x,y$. Typically they represent various linear combinations of the squares of internuclear distances between the three nuclei\cite{mead76} in a planar configuration. In this coordinate system the azimuthal angle $\phi$ is called the pseudorotation and $\rho=\sqrt{x^{2}+y^{2}} $ measures distortions from an equilateral triangle
configuration of nuclei. We describe the system by an amplitude $\psi(x,y,{\bm r})$ where ${\bm r}$ are electronic coordinates.  If the electronic, or fast, coordinates are integrated out one can approximate the adiabatic, or electronic, Hamiltonian as a truncated two-dimensional Hilbert space operator, which in the vicinity of the intersection is given by\cite{teller}
\beq
H_{ad}=
\left(
\begin{array}{cc}
  x  &  y    \\
  y  &  -x   \\
\end{array}
\right).
\label{S6.1}
\eeq
The eigenvalues of $H_{ad}$ are $\pm \sqrt{x^{2}+y^{2}}$ and correspond to first excited and ground states, respectively,
of the electronic Hamiltonian.  We approximate the vibronic kinetic energy operator
$ H_{KE} =-\frac{\hbar^{2}}{2 \mu} \Bigl( \frac{\partial^{2}}{\partial x^{2}} + \frac{\partial^{2}}{\partial y^{2}} \Bigr ) $ where 
$\mu$ is a reduced atomic mass.
Therefore, 
\beq
H = -\frac{\hbar^{2}}{2 \mu} \Bigl( \frac{\partial^{2}}{\partial x^{2}} + \frac{\partial^{2}}{\partial y^{2}} \Bigr ) + H_{ad}
\label{S6.2}
\eeq 
and has the form given by Eq. (\ref{z0.1a}).

Because $H_{ad}$ is real, Longuet-Higgins and Herzberg\cite{Herzberg} constrained its eigenstates
 to be real-valued and found
\beq
&& |\Phi_{g}  \rangle = \left ( \begin{array}{c}    
                            - \sin\phi/2 \\
                             \cos\phi/2 
                             \end{array}  \right )     ={\tilde U}(\phi) |g \rangle  \nonumber \\
&& {\tilde U}(\phi)=  \left ( \begin{array}{cc}    
                            \cos\phi/2  & -\sin\phi/2 \\
                            \sin\phi/2  & \cos\phi/2 
                             \end{array}  \right )   \quad                          
 |g \rangle = \left ( \begin{array}{c}    
                            0 \\
                             1
                             \end{array}  \right ), 
 \label{S6.2a}                             
\eeq
where $|\Phi_{g} \rangle $ is the ground adiabatic electronic state.
They noted that it is multivalued, 
as its value changes sign in traversing a circuit from $\phi=0$ to $\phi=2\pi$. The total system amplitude $\psi$
must be single valued and so in a Born-Oppenheimer approximation in which $  \psi= F(x,y) |\Phi_{g} \rangle $, the vibronic amplitude $F(x,y)$ must undergo a compensating sign change. That argument was used by Mead and Truhlar to invoke the minimal
coupling of the vibronic motion, in Eq. (\ref{S6.2}), with a vector potential given by
\beq
{\bm A}_{MAB} = \frac{\hat {\bm \phi}}{2 \rho}.   
\label{S6.3}
\eeq
In order to relate this result with the analysis in the previous section we offer a different tack. Instead of
constraining the eigenstates of Eq. (\ref{S6.1}) to be real we do not impose phase restrictions on them\cite{zyg87a}, but we do
require them to be single valued in parameter space $x,y$. We find
that 
\beq
&& H_{ad} = U_{c}(\phi) H_{BO} U_{c}^{\dag}(\phi)  \nonumber \\
\nonumber \\
&& H_{BO} = \left ( \begin{array}{cc}    
                            \sqrt{x^{2}+y^{2}} & 0  \\
                             0 &  -\sqrt{x^{2}+y^{2}}
                             \end{array}  \right )  
\label{S6.3a}
\eeq
where 
\beq
&& U_{c}(\phi) = \exp(-i \sigma_{2} \phi/2) \exp(-i \sigma_{3}\phi/2) = \nonumber \\
\nonumber \\
&&    \left(
\begin{array}{cc}
 e^{\frac{i \phi }{2}} \cos \left(\frac{\phi }{2}\right) & -e^{-\frac{i \phi
   }{2}} \sin \left(\frac{\phi }{2}\right) \\
 e^{\frac{i \phi }{2}} \sin \left(\frac{\phi }{2}\right) & e^{-\frac{i \phi
   }{2}} \cos \left(\frac{\phi }{2}\right) \\
\end{array}
\right).  
\label{S6.4}                        
\eeq
Unlike the operator $ {\tilde U}(\phi)$ given in Eq. (\ref{S6.2a}),  which undergoes a sign change as $\phi$ ranges
from $0$ to $2\pi$, $U_{c}(\phi)$  is single valued for all $\phi$, excluding the origin, i.e. $U(\phi+2\pi)=U(\phi)$. 
Repeating the derivations outlined in  the previous sections we arrive at the set of coupled
equations for the vibronic amplitudes in the adiabatic picture,
\beq
 -\, \frac{\hbar^{2}}{2 m} \Bigl ( {\bm \nabla} - i {\bm A} \Bigr )^{2} F({\bm R}) + 
{ { V}}_{BO}({\bm R}) F({\bm R}) = E F({\bm R}). \nonumber \\
\label{S6.5}
\eeq
$F({\bf R})$ is a column vector whose entries are the ground and excited state vibronic amplitudes,
 ${ V}_{BO}({\bm R} )$ is the BO eigen-energy matrix  and 
${\bm A}$ is the non-Abelian gauge  potential
\beq
&& {\bm A} = i U_{c}^{\dag} {\bm {\nabla}} U_{c} = 
\frac{1}{2\rho}  \left(
\begin{array}{cc}
   -1 &  -i \exp(-i \phi)   \\
    i \exp(i \phi)   &  1  
\end{array}
\right).
\label{S6.6}
\eeq
It is identical to expression (\ref{S2.2}) for the case $\Omega=3\pi/2$.
This equivalence is a consequence of the fact that
 $U(\phi)$, and $U_{c}$ are related, as shown in Appendix C, by a constant unitary matrix.  A Born-Oppenheimer projection of Eq. (\ref{S6.5})  leads to Eq. (\ref{z0.3}) where $ {\bm A}_{P}$ becomes the Mead-Truhlar gauge potential ${\bm A}_{MAB}$.

$U({\bm R})$ defined in Eq. (\ref{z1.1c}) diagonalizes the adiabatic Hamiltonian $ \mu_{B} {\bm \sigma}\cdot {\bm B} $, which
for the ${\bm B}$ configuration Eq. (\ref{z1.1a}) does not possess degeneracy in ${\bm R}$ space.  $ U_{c}$ diagonalizes
Hamiltonian (\ref{S6.1}) which exhibits a conical intersection at the origin of ${\bm R}$ space. Thus the emergence of the non-trivial gauge structures follow from the properties of the unitary operators that diagonalize
the respective  adiabatic Hamiltonians, and is insensitive to the fact that  $H_{BO} $ may possess degeneracy. 
In Appendix C we show how both $U({\bm R})$ and $U_{c}$ can be constructed, via a Wilson line integral, from their respective gauge fields.

\begin{figure}[ht]
\centering
\includegraphics[width=0.9 \linewidth]{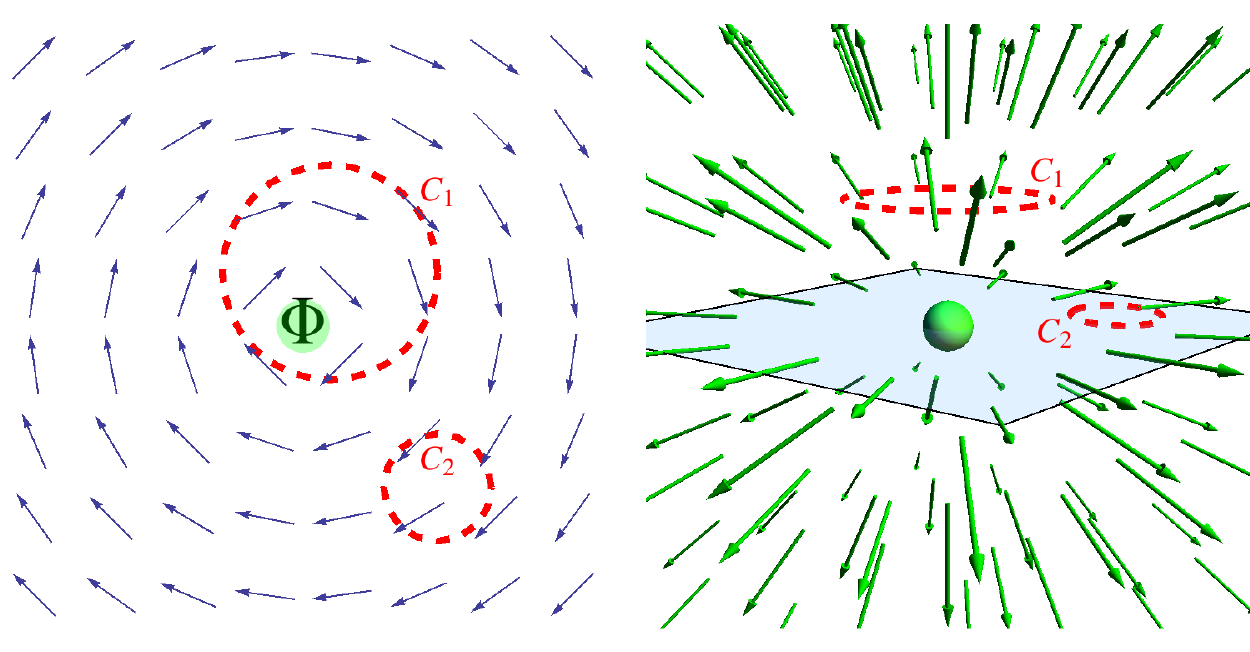}
\caption{\label{fig:fig14} (Color online). The left panel illustrates the vortex configuration, projected on the $xy$ plane, of the external magnetic field given by Eq. (\ref{z1.1a}), and the red dashed lines are paths in ${\bm R}$ space.
$\Phi$ represents the effective flux tube pointing into the page.
The right hand panel shows  field lines emanating from the effective Dirac monopole located at the origin. The dashed red lines represent  loops in the  parameter space of the monopole, and are maps of the corresponding  loops in ${\bm R}$ space. The blue background plane bisects the monopole.}
\end{figure}

Finally, we explore the relationship of this analysis with that given in discussions of Berry's phase\cite{ber84}. The adiabatic  Hamiltonian  $ H_{ad} = \mu_{B} {\bm \sigma} \cdot {\bm B}$, given in Eq. (\ref{z1.1b}),  shares the  structure of Berry's model. In the latter ${\bm B}$ is taken as a classical parameter that can be arbitrarily varied. Under adiabatic conditions,  i.e. slow variation of ${\bm B}$,  a spin $1/2$ system accrues a phase that is determined by the vector potential, or connection, of a magnetic monopole (see the right panel of Figure \ref{fig:fig14} ). The gauge field of the monopole differs from that given in Eq. (\ref{S2.4}) which describes 
a magnetic flux tube extended along the $z$ axis. In our discussion ${\bm B}$ is a fixed external magnetic field, shown by the vortex lines in the left panel of Figure \ref{fig:fig14}, and gauge field Eqs. (\ref{S2.2},\ref{S2.4}) follows from it by considering variation of parameter ${\bm R}$ instead of ${\bm B}$. The gauge structure induced by the former, realized in molecular systems with conical intersections, does not lead to a Lorentz force. However,  effective magnetic monopoles in ${\bm R}$ space are realized in diatoms\cite{moo86,zyg87a,zyg90}. They lead to effective Lorentz
forces, (i.e. geometric magnetism), in those systems\cite{zyg87a,zyg90}. In addition, ${\bm R}$ is a quantum variable and so true adiabaticity  is ill defined. Nevertheless, if we demote ${\bm R}$ to a c-number we can consider loops in ${\bm R}$ space and relate them to corresponding loops in ${\bm B}$ space. This is illustrated in Figure (\ref{fig:fig14}). In the left panel of this
figure we show a loop $C_{1}$ which encloses the flux tube, of magnitue $\Phi$, at the origin. In ${\bm B} $ space it corresponds to
a loop that subtends a solid angle $\Omega$, from the location of the monopole,  so that $ \Phi = \pi(1- \cos \Omega)/2 $. A loop
$C_{2}$ that does not enclose the flux tube corresponds to a loop in ${\bm B}$ space that lies in a plane (shown by the blue background in the right panel of that figure) that bisects the monopole.

\section{Summary and Discussion}
These examples  illustrate how induced geometric phase gauge potentials, that lead to holonomies and effective ``magnetic'' forces,  arise in simple quantum systems without
resorting to assumptions that are based on  BO - type approximations.  We introduced a two-level Hamiltonian which mimics the behavior of a charged particle
that is subjected to both electric and magnetic fields in such a way that enable a velocity selector.
In another example, that of a neutral spin 1/2 system subjected to an external magnetic field, we showed how AB-like effects arise
in low energy solutions to the coupled Schroedinger equations.   We introduced and discussed both an interferometry setup, as well as a scattering  scenario, in which such effects arise.
In addition we showed how the system transitions from that described by an Abelian gauge potential to that in which non-Abelian effects manifest at higher collision energies. 

Dirac\cite{Dirac31} noted that if a solution $\psi$ to the Schroedinger equation is multiplied by a phase factor
so that $ \psi({\bm R}) \rightarrow \psi' ({\bm R}) = \exp(i \alpha({\bm R})) \, \psi({\bm R}) $,
and since we can always represent 
$ \alpha({\bm R}) = \int^{\bm R}_{C}  d{\bm r}\cdot {\bm A}({\bm r} )$
this substitution leads to the replacement $ {\bm \nabla} \psi \rightarrow {\bm \nabla}\psi' 
- i {\bm A} \psi'. $  However, conventional quantum mechanics demands that $\psi'({\bm R})$ be single-valued at all ${\bm R}$, and this condition constrains the gauge potential ${\bm A}$ to a trivial, pure, gauge which does not lead to new physics. Dirac suggested that minimal coupling of a charged particle with a non-trivial gauge field 
follows from a non-integrable factor $\alpha({\bm R,C})$, a functional that depends on the path $C$. There have been explorations in that direction\cite{stam62} but difficulties in enforcing single-valuedness has limited the utility of this approach.

The examples provided above illustrate how an aspect of Dirac's program is realized. Let $\psi$ be a multi-component amplitude and consider the non-Abelian version of Dirac's substitution, i.e. $ \psi \rightarrow \psi'= U({\bm R}) \psi $, where $U({\bm R})$ is a 
single-valued, differentiable unitary matrix operator. In \cite{zyg12} and in Appendix C we showed how $U({\bm R})$ can be expressed as a Wilson line of a non-Abelian, pure,
gauge potential ${\bm A}$. Because ${\bm A}$ is a pure gauge, issues involving multi-valuedness in $U({\bm R})$ and hence in $\psi'$ do not arise. Nevertheless, as our examples  demonstrate, if the gauge symmetry is broken by  energy gaps (i.e. non-degeneracy of $H_{BO}$ )  non-trivial gauge fields that lead to gauge forces and/or topological holonomies can emerge in low energy solutions to the Schroedinger equation.    

\appendix
\section{Packet dynamics}
At lower collision energies, in which the excited Zeeman level is closed, we consider the propagation
of a coherent wave packet that is initially localized, as shown in Figure (\ref{fig:fig2}). Using the dimensionless
coordinates defined in \cite{zyg12}, $\xi=x/L,\eta=y/L,\tau= \frac{\hbar}{2m L^{2}} \,  t $,
we take the free particle wave packet (normalized to the value 2)
\beq
&& \psi(\xi,\eta,\tau) = \psi_{1}(\xi,\eta,\tau) + \psi_{2}(\xi,\eta,t) \nonumber \\
&& \psi_{i}= \int \int \, d{k}_{1}  d{k}_{2}
\Phi({k}_{1}- {k}_{x_{i}} ) \Phi({k}_{2}- {k}_{y_{i}}) \times \nonumber \\
&& \exp(- i \tau ( k^{2}_{1} + k^{2}_{2} ) )  \, \exp(-i \tau {\tilde \Delta} )  \times \nonumber \\
&& \exp(i k_{1}(\xi-\xi_{i})) \exp(i k_{2}(\eta-\eta_{i}) ).
\nonumber \\
\label{S3.1}
\eeq
With the choice 
\beq
\Phi(k) = \frac{\sqrt{a}}{{(2 \pi^3)}^{\frac{1}{4}}} \exp(-a^2 k^2)
\label{S3.2}
\eeq
$\psi_{i}(\xi,\eta,\tau)$ is a solution to the equation
\beq
i \frac{\partial \psi_{i}}{\partial \tau} = - \frac{\partial^{2} \psi_{i}}{\partial \xi^{2} } 
 - \frac{\partial^{2} \psi_{i}}{\partial \eta^{2} } + \tilde {\Delta} \psi_{i} 
\label{S3.3}
\eeq
and describes, at $\tau=0$, a Gaussian wavepacket of width $a$ localized at $(\xi_{i},\eta_{i})$ with momenta
$ k_{x_{i}}, k_{y_{i}} $. $\tilde \Delta$ is the energy defect $\Delta$ in units of $ \frac{2 m L^{2}}{\hbar^{2}}$
 By linearity, the coherent sum $ \psi(\xi,\eta,\tau)$ is also a possible
packet.
Here we chose $ \xi_{1}= \xi_{2}=0, \eta_{1}=-\eta_{0}, \eta_{2}=\eta_{0}  $,
and $ k_{x_{1}} =k_{x_{2}} =k,  k_{y_{1}} = - k_{y_{2}} =k $. 
At time $ \tau_{c}= \eta_{0} /2k $ the two packets coalesce and form 
the interference pattern illustrated in Figure \ref{fig:fig2}. The center of the merged packets
at  $\tau_{c} $ passes through the line $\xi = \eta_{0} $. On it 
\beq
&& \psi(\eta_{0},\eta,\tau_{c} ) =  \int \int \, d{k}_{1}  d{k}_{2} \, 
\Phi({k}_{1}- k ) \Phi({k}_{2}- k )  \times \nonumber \\
&& \exp(- i \tau_{c} ( k^{2}_{1} + k^{2}_{2} ) ) \exp(i k_{1} \eta_{0}) 
\exp(- i \tau {\tilde \Delta} )
\times  \nonumber \\
 && \Bigl [ \exp(i k_{2} (\eta+\eta_{0})) + \exp(-i k_{2} (\eta- \eta_{0}))  \Bigr ].
\label{S3.4}
\eeq
Integration of Eq. (\ref{S3.4}) yields\beq
&& |\psi(\eta_{0},\eta,\tau_{c}) |^{2} = \nonumber \\
&& \frac{8 a^{2} k^{2} \exp(-\frac{ 2 a^{2} k^{2} \eta^{2}}{4 a^{4} k^{2} + \eta_{0}^{2}})}
{ 4 a^{4} k^{2} \pi + \pi \eta_{0}^{2} } \cos^{2}(k \eta). 
\label{S3.5}
\eeq

\section{ AB scattering wave function in the presence of an impenetrable cylinder}
Scattering by an impenetrable cylinder, in the presence of a AB flux tube, has been extensively reviewed in the literature
\cite{Au84,Berry80,march92}. Below we provide an analysis following along lines given in Ref. \cite{Berry80}.
We consider  the Schroedinger equation,
\beq
&& \Bigl ( {\bm {\nabla}} - i {\bm A} \Bigr )^{2}\psi + k^{2} \psi - U(r) \psi =0 \nonumber \\
&& {\bm A} = {\hat {\bm \phi}} \frac{\Phi}{2 \pi r}
\label{1.0}
\eeq
where ${\bm A}$ is a vector potential, and $U(r)$ is a short range potential, wich we
take to be that of an  impenetrable  cylinder whose axis is perpendicular to the incoming flux.
Expressing the wave function
\beq
\psi(r,\phi)=\sum_{m=-\infty}^{\infty}  c_{m} \exp(i m \phi) \, R_{m}(r)
\label{1.1}
\eeq
we obtain the radial equation,
\beq
&& R''(r) + \frac{R'(r)}{r}-\frac{(m-\alpha)^{2}}{r^{2}} R(r) +  \nonumber \\
&& (k^{2}  - U(r)) R(r)=0
\label{1.2}
\eeq
where $ \alpha \equiv \Phi/2\pi $.
Its solutions
are linear combinations of
\beq  J_{|m-\alpha|}(k r) \quad N_{|m-\alpha|}(k r). \nonumber \\
\nonumber \eeq
For the repulsive core we require $R(a)=0$ and therefore the radial
solution has the form
\beq
R(r) = c_{1} J_{|m-\alpha|}(k r)+ c_{2} H^{(1)}_{|m-\alpha|}(k r) 
\label{1.4}
\eeq
where 
\beq
\frac{c_{2}}{c_{1}}=-\frac{J_{|m-\alpha|}(k a)}{H^{(1)}_{|m-\alpha|}(k a)}.
\label{1.5}
\eeq
We define the AB amplitude (\cite{ab59})
\beq
\psi_{ab}(r,\phi)= \sum_{m=-\infty}^{m=\infty} (-i)^{|m-\alpha|} J_{|m-\alpha|}(k r) \exp(i m \phi),
\nonumber \\
\label{1.6}
\eeq
and so choose $ c_{1}= (-i)^{|m-\alpha|} $.  Therefore, according to Eq. (\ref{1.5}) 
\beq
c_{2} = - (-i)^{|m-\alpha|} \frac{J_{|m-\alpha|}(k a)}{H^{(1)}_{|m-\alpha|}(k a)}.
\label{1.8}
\eeq
and
\beq
&& \psi(r,\phi) = \psi_{ab}(r,\phi) - \psi_{R}(r,\phi) \nonumber \\
&& \psi_{R}(r,\phi)  \equiv   \sum_{m=-\infty}^{\infty} 
 -i^{|m-\alpha|}  \frac{J_{|m-\alpha|}(k a)}{H^{(1)}_{|m-\alpha|}(k a)} \, {H^{(1)}_{|m-\alpha|}(k r).} \nonumber \\
 \label{1.9}
 \eeq
For the value $\alpha =1/2$, the so called maximal AB scattering case, $\psi_{ab}(r,\phi)$, is given 
by the closed-form expression \cite{ab59,Berry80}
\beq
&& \psi_{ab}(r,\phi) = \frac{2}{\sqrt{\pi}}  \exp(i \phi/2) \exp(-i k z )
\exp(-i\pi/4) \times \nonumber \\ 
&& \exp(\mp i \pi/4)  Erf(\exp(\mp i 3\pi/4) \sqrt{2 k r} \cos(\phi/2) )
\label{1.10}
\eeq
where the $\pm$ refers to the regions $  0 <  \phi \leq  \pi$ and $ -\pi < \phi \leq 0 $ respectively. 
\section{Representation of a unitary matrix by a pure gauge field}
Consider a unitary matrix $ U ({\bm R}) $ that is continuous, single valued, and has a derivative defined at each point ${\bm R}$, excluding the origin, so
that
\beq
{\bm A}({\bm R}) = i\, U^{\dag}({\bm R}) \, {\bm \nabla} U({\bm R}) 
\label{2.1}
\eeq
is a pure gauge potential\cite{wuyang75}. Suppose it has the
property
\beq
P \exp(i  \oint  {\bm dR} \cdot {\bm A}) =1 
\label{2.1a}
\eeq
for any closed path in ${\bm R}$ space. Can we then represent $U({\bf R})$ as a Wilson line of the corresponding
gauge potential ? 
Here we show that this is the case for the operators $U({\bm R})$  given in  Eq.(\ref{z1.1c}),  and  $U_{c}$ defined in Eq. (\ref{S6.4}).
We define a path $C$ so that ${\bm R}(t) = R_{0} \cos(\omega t) {\bm {\hat i}} + R_{0} \sin(\omega t) {\bm {\hat j}} $ on it, and traverses a loop of radius $R_{0}$ in the counterclockwise direction. Consider the (Wilson) line integral along this
path from $\phi_{0}$ to $\phi$
\beq
&& W(\phi,\phi_{0},C) = P \exp(i \int_{C} d {\bm R} \cdot {\bm A}({\bm R}))  \equiv  \nonumber \\ 
&& W(t,t_{0}) = T  \exp(i \int_{t_{0}}^{t}  A(t) ) \nonumber \\
\label{2.2}
\eeq
where, $ \phi/\omega=t, \phi_{0}/\omega = t_{0} $, $T$ is the Dyson time -ordering operator,
\beq
&& A(t) \equiv    \frac{d {\bm R}(t)}{d t} \cdot { {\bm A}({\bm R}(t)}) = \nonumber \\
&& \left(
\begin{array}{cc}
  -\omega  \sin
   ^2\left(\frac{\Omega }{2}\right) & \frac{1}{2} i e^{-i t \omega } \omega  \sin (\Omega ) \\
 -\frac{1}{2} i e^{i t \omega } \omega  \sin (\Omega ) & \omega  \sin
   ^2\left(\frac{\Omega }{2}\right) \\
\end{array}
\right)
\label{2.3}
\eeq 
and we used Eq. (\ref{S2.2}). Noting that
\beq
\frac{ dW(t)}{dt} = i A(t) W(t) 
\label{2.4}
\eeq
we integrate to get
\begin{widetext}
 \beq
 W(\phi,\phi_{0}) = U^{\dag}(\phi) \,  U(\phi_{0}) = 
 \left(
\begin{array}{cc} \cos ^2\left(\frac{\Omega }{2}\right)+\sin ^2\left(\frac{\Omega }{2}\right)
   e^{-i (\phi -\phi_{0} )}
  & -e^{-i \frac{\phi+\phi_{0}}{2}  }  \sin(\frac{\phi-\phi_{0}}{2}) \sin(\Omega) \\
 e^{i \frac{\phi+\phi_{0}}{2}  }  \sin(\frac{\phi-\phi_{0}}{2}) \sin(\Omega)
 & \cos ^2\left(\frac{\Omega }{2}\right)+\sin ^2\left(\frac{\Omega }{2}\right)
   e^{i (\phi -\phi_{0} )} \\
\end{array}
\right).
\label{2.5}
 \eeq
 \end{widetext}
$W(\phi,\phi_{0}) $ obeys the group property
\beq
W(\phi,\phi_{0}) = W(\phi,\phi_{1}) W(\phi_{1},\phi_{0}) 
\label{2.6}
\eeq
for $ \phi > \phi_{1} >\phi_{0} $, and since $W$ is unitary we have
\beq
&& W^{\dag}(t,t_{0}) = {\tilde T} \exp(-i \int_{t_{0}}^{t} A(t)) = {\tilde T} \exp(i \int_{t_{0}}^{t} {\tilde A}(t)) \nonumber \\
&& {\tilde A} (t) =   \frac{d {\bm {\tilde R}}(t)}{d t} \cdot { {\bm A}({\bm {\tilde R}}(t)}) 
\label{2.7}
\eeq
where $ {\tilde T}$ is the anti-time ordering operator, and $ {\tilde {\bm R}}(t)$ defines a circular path running along
the clockwise direction.
According to Eq. (\ref{2.5})
\beq
 U(\phi) = U(\phi_{0})W^{\dag} (\phi,\phi_{0})  
\label{2.8}
\eeq
$$ $$
where $U(\phi)$ is given in Eq. (\ref{z1.1c}). 
\begin{figure}[ht]
\centering
\includegraphics[width=0.5 \linewidth]{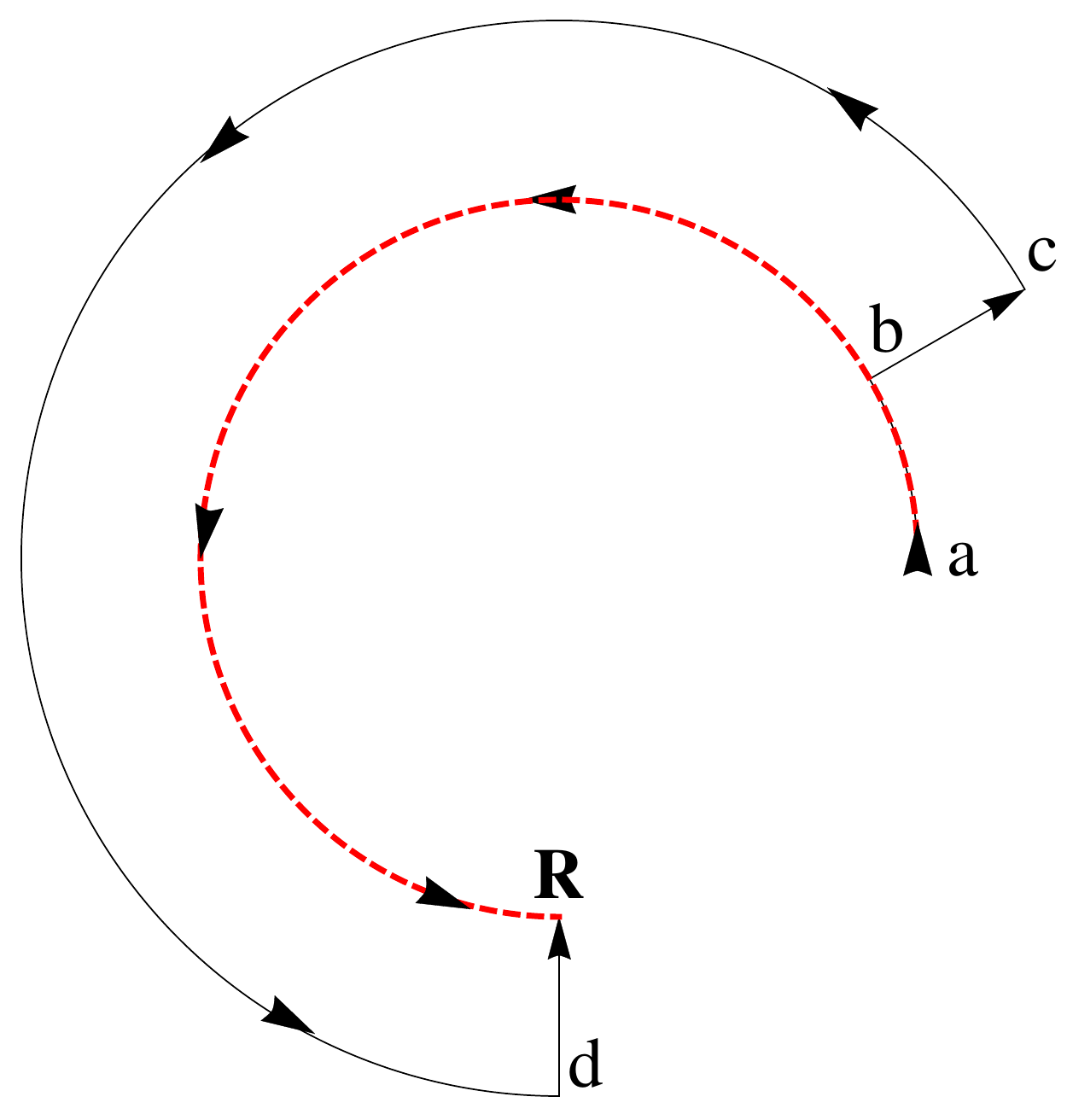}
\caption{\label{fig:fig15} (Color online).
Integration paths for Wilson line integral, where point $a$ lies on a ray $\phi=\phi_{0}$ and ${\bm R}$ lies
on the ray $\phi$. }
\end{figure}
Therefore we have shown that $U({\bm R})$ can be defined
in terms of a line integral along the path shown by the red dashed arc in Fig. (\ref{fig:fig15}). 
Point $a$ is along the line
$\phi_{0}$ (not including the origin)  and $ {\bm R}$ is some arbitrary point (not on the origin). 
However we can deform the path as shown
by the solid line segments in that figure. By the group property Eq. (\ref{2.6}) we can equate 
\beq
W({\bm R},\phi_{0}) = W({\bm R},d) W(d,c)W(c,b)W(b,a)
\label{2.9}
\eeq
where we used $W(c,b)=W({\bm R},d) =1$,
and demonstrate that $U({\bm R})$ can be expressed in terms of
the Wilson line, for any right handed path, provided that it does not
pass through the origin. We stress, again, such a representation is not possible for the Abelian projection
\beq
{\bm A}_{P}  \equiv  Tr \, P {\bm A} P =  {\bm {\hat \phi}} \frac{1 - \cos\Omega}{2 \rho} 
\label{2.10}
\eeq
as its Wilson loop integral, for any loop surrounding the origin, is not unity. As a consequence
$$ U_{P}  \equiv  \exp(i \, \int_{C} d{\bm R}\cdot {\bm A}_{P} ) $$ is multivalued.  In the same
way we also show that operator  $U_{c}(\phi)$, defined in Eq. (\ref{S6.4}) is given
by
\beq
U_{c}(\phi) = W^{\dag}(\phi,0)
\label{2.11}
\eeq
where we have set  $\Omega= 3\pi/2$ in the expression for $W$. Comparing Eq. (\ref{2.11}), setting
$ \Omega=3 \pi/2 $,  with Eq. (\ref{2.8}) we find that
\beq
&& U(\phi) = U(0) U_{c}(\phi)  \nonumber \\
&& U(0) = \exp(3\pi i/2 \, \sigma_{1} ).
\label{2.12}
\eeq
\begin{acknowledgments}
I  thank Dr. M. Kiffner, and an anonymous reviewer, for useful comments and suggestions.
\end{acknowledgments}


\begin{thebibliography}{40}
\bibitem [{com()}]{comment}%
  \BibitemOpen
  \href@noop {} {}\bibinfo {note} {If, as in Berry's Hamiltonian, the adiabatic
  energies are degenerate at an isolated point in a 3-dimensional parameter
  space, $U$ can be made to be single valued, in a region not including this
  point, provided that it is defined as two overlapping charts, i.e. $U({\bm
  R})=\exp(-i \phi/2 \sigma_{3}) \exp(-i \theta/2 \sigma_{2}) \exp(\pm i \phi/2
  \sigma_{3}) $. See also Ref. \cite{zyg90}.}\BibitemShut {Stop}%
\end{thebibliography}

%

\end{document}